\documentclass[aps,amsfonts,pra,twocolumn,showpacs]{revtex4}
\usepackage{epsfig,amsmath,amssymb,bm,epsf,graphicx,psfrag}
\usepackage[all]{xy}
\usepackage{graphicx}
\usepackage{color}

\pdfoutput=1

\newtheorem{Theorem}{Theorem}

\newtheorem{Cor}{Corollary}

\newtheorem{Lemma}{Lemma} 

\newtheorem{Def}{Definition}

\pdfoutput=1

\begin{document}

\title{{Contextuality and Wigner function negativity in qubit quantum computation}}

\author{$\text{Robert Raussendorf}^1$, $\text{Dan E. Browne}^2$, $\text{Nicolas Delfosse}^{3,4,5}$, $\text{Cihan Okay}^6$, $\text{Juan Bermejo-Vega}^{7,8}$\\ \mbox{ }}

\affiliation{1: Department of Physics and Astronomy, University of British
Columbia, Vancouver, BC, Canada,\\
2: Department of Physics and Astronomy, University College London, Gower Street, London, UK,\\
3: D{\'e}partement de Physique, Universit{\'e} de Sherbrooke, Sherbrooke, Qu{\'e}bec, Canada,\\ 
4: IQIM, California Institute of Technology, Pasadena, CA, USA,\\
5: Department of Physics and Astronomy, University of California, Riverside, California, 92521, USA,\\
6: Department of Mathematics, University of Western Ontario, London, Ontario, Canada,\\
7: Max-Planck Institute for Quantum Optics, Theory Division, Garching, Germany,\\
8: Dahlem Center for Complex Quantum Systems, Freie Universit{\"a}t Berlin, Berlin, Germany
}

\date{\today}

\begin{abstract} We describe schemes of quantum computation with magic states on qubits for which contextuality and negativity of the Wigner function are necessary resources possessed by the magic states. These schemes satisfy a constraint. Namely, the non-negativity of Wigner functions must be preserved under all available measurement operations. Furthermore, we identify stringent consistency conditions on such computational schemes, revealing the general structure by which negativity of Wigner functions, hardness of classical simulation of the computation, and contextuality are connected.
\end{abstract}


\maketitle

\section{Introduction}\label{Intro}

Contextuality \cite{Bell}--\cite{Gueh} and negativity of Wigner functions \cite{Wigner}--\cite{Contin} have recently  been established as a necessary resources for quantum computation by injection of magic states (QCSI) \cite{Howard}--\cite{Galv2}. This was first achieved for systems of qudits \cite{NegWi}, \cite{Howard}, where the local degrees of freedom have an odd number of states. These findings beg the question: ``Under which conditions are contextuality and Wigner function negativity computational resources for QCSI on $n$-qubit systems?''.

This is the question we address in the present  paper. As a first step into this direction, it was shown in \cite{ReWi} that Wigner function negativity and contextuality of magic states are necessary resources for QCSI on rebits, which are systems of local dimension 2 whose state vector is real-valued. Here we generalize that approach.

Before we state our results, we give a brief description of the technical terms that have appeared above.
{\em{Quantum computation with magic states}} (QCSI) \cite{BK} is closely related to the standard circuit model. It deviates in that the allowed state preparations, unitary transformations and measurements are restricted to be non-universal and, in fact, efficiently classically simulable. These operations are called ``free''. Computational universality is restored by the capability to inject so-called magic states, which are states that cannot be prepared by the free operations. The source of computational power thus shifts from the gates to the magic states, and it is natural to ask which quantum features must be present in these states to allow for quantum speedup.

{\em{Wigner functions}} \cite{Wigner}--\cite{Mari} describe quantum states in phase space. They are quasi-probability distributions, and as such the closest quantum analogue to joint probability distributions of position and momentum in classical statistical mechanics. The difference is that Wigner functions can take negative values, and this negativity is a signature of quantumness \cite{Hudson}, \cite{Spekk}.

{\em{Contextuality}} means that measurements cannot be viewed as deterministically revealing pre-existing properties of a system. This happens in Hilbert spaces of dimension $d\geq 3$ \cite{KS}. There, it is impossible to consistently assign pre-existing values $\lambda$ to all observables, in such a way that $\lambda(A)$ is a property of the observable $A$ alone, and not also of compatible observables that are measured simultaneously. 

State-dependent contextuality is a weaker form that can be attributed to quantum states rather than observables \cite{Abram}, and it is the notion we are interested in here. The set of measurable observables may be restricted in such a way that consistent non-contextual value assignments exist. Then, a probabilistic mixture of such value assignments is a non-contextual hidden variable model (ncHVM). For certain quantum states, it may happen that no probability distribution over the value assignments correctly reproduces all measurement statistics. Quantum states of this kind are called contextual. 

The phenomenon of state-independent contextuality wrt. Pauli observables (which are the free observables in QCSI) represents an obstacle towards establishing contextuality of the magic states as a computational resource. Namely, because of it, for $n\geq 2$ qubits, there does not even exist a single consistent non-contextual value assignment for the Pauli observables. Hence no quantum state of $n\geq 2$ qubits, not even the completely mixed state, can be described by a ncHVM. Every quantum state is thus contextual. And if contextuality is ubiquitous, it cannot be a resource.
\smallskip

In this paper, we overcome the problem posed by state-independent contextuality, by restricting to QCSI schemes on qubits which satisfy an additional condition. Under this condition, contextuality and Wigner function negativity of the magic states are necessary computational resources for QCSI on qubits. The condition is
\begin{itemize}
 \item[(P1)]{Non-negative Wigner functions remain non-negative under free measurements.} 
 \end{itemize}
Condition (P1) is the very basis for the usefulness of  Wigner functions in the description of QCSI, namely to reveal the near-classicality of QCSI without the magic states. This property also holds for the previously discussed cases of qudits and rebits. (P1) restricts the Pauli observables that are available fore measurement. To motivate this constraint, we observe that many constructions for fault-tolerant quantum computation are QCSI schemes in which the free gates are a strict subset of the Clifford  gates. One example is the first concrete QCSI scheme ever proposed, with the free operations provided by braiding and fusion of Ising anyons, and magic states enabling CNOT and $\pi/8$-gates \cite{IsA}.

However, (P1) is trickier than might at first appear. For a start, we do not require a counterpart of (P1) for the unitary operations available in QCSI, and imposing it would indeed be too restrictive. Those unitaries may introduce large amounts of negativity into the Wigner function without compromising efficient classical simulability.

Condition (P1) can be used to define the free measurements in a QCSI scheme, given a Wigner function. Typically, but not always, the resulting set of free measurements is small. Thus the question arises whether the corresponding QCSI schemes really operate on the intended number of qubits. To filter out the ``true'' $n$-qubit QCSI schemes, we  therefore impose a second condition.
\begin{itemize}
 \item[(P2)]{The available measurements are tomographically complete.} 
\end{itemize} 
Condition (P2) means that with the free operations of an $n$-qubit QCSI scheme the density matrix $\rho$ of any $n$-qubit quantum state can be fully measured. Rebit QCSI \cite{ReWi}, for example, does not satisfy it. 

Conditions (P1) and (P2) are hard to satisfy jointly. Informally speaking, for typical Wigner functions, (P1) generates small sets of free measurements whereas (P2) demands large sets. Yet, sets of freely measurable observables that satisfy both (P1) and (P2) do exist for every number $n$ of qubits.\medskip 

{\em{Results.}} We show that for all $n$-qubit QCSI schemes satisfying (P1), contextuality and Wigner function negativity are necessary for quantum computational universality and quantum speedup. The result on contextuality is strictly stronger than our corresponding result on Wigner functions. Different from the qudit case \cite{Howard}, \cite{Nicol}, magic states can have negative Wigner functions but still be non-contextual. Consequently, the fundamental notion of classicality for qubit QCSI schemes subject to (P1) is a non-contextual hidden variable model, not a positive Wigner function. 

Our result on contextuality is an extension of the corresponding results for qudits \cite{Howard}, where the Hilbert space dimension of the local systems is an odd prime or a power of an odd prime, and for rebits \cite{ReWi}, where the Hilbert space dimension of the local systems is 2, but the density matrix is constrained to be real. 

Our result on Wigner-negativity-as-a-resource extends \cite{NegWi} from qudits to qubits, and is complementary to the earlier work \cite{Galv2} on qubits. In \cite{Galv2}, many Wigner functions are simultaneously used to characterize the classical region for the magic states, whereas our work only involves one. On the other hand, the present  result only applies for QCSI schemes satisfy the condition (P1), whereas \cite{Galv2} applies in general. For QCSI schemes where both methods can be applied, the present result labels a larger region of the state space as ``guaranteed classical''.

\section{Motivation, results and outline}

In this section we give a broader motivation for our work, explain why the qubit case requires a separate discussion, summarize our results, and provide an outline of the remainder of the paper.

\subsection{Motivation}\label{2diffA}

Which quantum property makes the magic states valuable for QCSI? This is the question we address in the present paper, and to approach it, we first need to clarify in which sense the free operations of QCSI by themselves are not fully quantum and not computationally valuable. These operations are certainly not entirely classical. For example, highly entangled states can be created by them. 

In the present paper, the term ``qudit'' always refers to systems of odd Hilbert space dimension. In this benign case of qudits, the near-classicality of the free operations is revealed by the following characteristics:  The free operations (which are Clifford operations) preserve the properties of quantum states (i) to be stabilizer states, and hence to be efficiently classically simulable by the stabilizer formalism \cite{Stab}, (ii) to have a non-negative Wigner function, and (iii) to be describable by a non-contextual hidden variable model. 

Thus there are different angles from which to view classicality in QCSI---the quantum optics angle, focussing on Wigner functions; the quantum foundations angle, focussing on ncHVMs, and the stabilizer angle, focussing on efficient classical simulability. Remarkably, in the qudit case, those different angles amount to essentially the same. First, the notions of contextuality and Wigner function negativity agree for qudits \cite{Howard}, \cite{Nicol}, for the Wigner function proposed in \cite{Gibbons}, \cite{Gross}. Next, if the magic states have a non-negative Wigner function, the resulting QCSIs can be efficiently classically simulated \cite{NegWi}. The domain of applicability of this simulation method strictly contains the domain of applicability of the stabilizer method.

When transitioning from qudits to qubits, this well-rounded picture of classicality splinters. The stabilizer simulation method survives unharmed. Wigner functions can still be defined, although their definition cannot be straightforwardly adapted from the infinite-dimensional or finite odd-dimensional case \cite{Mari}, \cite{Zhu1}, \cite{Zhu2}. Their non-negativity is no longer preserved under all Clifford gates. 

Worst fares contextuality of the magic states. As already noted in the introduction, due to the phenomenon of state-independent contextuality with Pauli-observables \cite{Merm}, ncHVMs can no longer be consistently defined. Hence every quantum states on $n\geq 2$ becomes contextual \cite{Howard}, and contextuality of the magic states is no longer a resource. 

The latter represents a severe obstacle to extending the result \cite{Howard} to qubits. We overcome it by invoking the additional assumption (P1). The phenomenon of state-independent contextuality with Pauli observables most pronouncedly illustrates that qubits are different.

\subsection{Why are qubits different?}\label{2diff}

The difference between qudits and qubits derives from the fact that Heisenberg-Weyl operators behave differently---in ways that matter for contextuality and Wigner functions---depending on whether the Hilbert space dimension is even or odd. The HW operators are important for QCSI because, by construction, all freely measurable observables are of this type. 

Denote by $d$ the dimension of the local Hilbert space, and by $\textbf{x}$ and $\textbf{z}$ two HW operators acting on that space,
\begin{equation}\label{LHWdef}
\textbf{x} := \sum_{i=0}^{d-1} | i + 1\, \text{mod}\, d\rangle \langle i|, \;\textbf{z} := \sum_{i=0}^{d-1} \omega^i |i\rangle \langle i|,
\end{equation}
where $\omega = e^{2\pi i/d}$.  The definition of the HW operator $\textbf{x}$ in Eq.~(\ref{LHWdef}) requires addition mod $d$, and in result, addition mod $d$ is a common operation when reasoning about Heisenberg-Weyl operators in dimension $d$. What distinguishes the qubit from the qudit case is that $2^{-1} \mod d$ is well-defined if $d$ is odd, i.e., there exists an $x \in \mathbb{Z}_d$ such that $2x = 1 \mod d$. But $2^{-1} \mod d$ is undefined if $d$ is even, specifically if $d=2$.  

Let's see how this mathematical difference is relevant for physics. First, we remark that the non-existence of the inverse of $2$ (mod $d$) affects the adaption of Wigner functions from the infinite-dimensional to the finite-dimensional case. The standard construction \cite{Gross}, \cite{Mari} works only for odd $d$, precisely for the above reason. 

Here, we discuss the existence of state-independent proofs of contextuality based on HW operators (HW-SIC), such as the Peres-Mermin square \cite{Merm}, \cite{AP} and Mermin star \cite{Merm}, in tensor product Hilbert spaces of dimension $d^n$. As we discussed in the introduction,  if state-independent contextuality is present, then contextuality cannot be a resource possessed only by some magic states.  

The Heisenberg-Weyl operators $\tau^\gamma_\textbf{a}$ in the Hilbert space of $n$ systems of dimension $d$ are
$$
\tau^\gamma_\textbf{a} := \omega^{\gamma(\textbf{a})}Z(\textbf{a}_Z)X(\textbf{a}_X),
$$
where the function $\gamma$ is an arbitrary phase convention, $\textbf{a}=(\textbf{a}_Z,\textbf{a}_X)$, and $Z(\textbf{a}_Z) = \bigotimes_{i=1}^n ( \textbf{z}^{(i)})^{a_{Z,i}}$. 

There is no HW-SIC if $d$ is odd. First note that the existence or non-existence of HW-SIC is independent of the choice of $\gamma$. We may thus choose a convenient such function, which for our purposes is
\begin{equation}
\gamma_0(\textbf{a}) := \textbf{a}_Z\textbf{a}_X/2 \mod d.
\end{equation} 
This requires the existence of $2^{-1} \mod d$, and thus only works if $d$ is odd.
By direct calculation we find that
$$
\tau_\textbf{a}^{\gamma_0} \tau_\textbf{b}^{\gamma_0}=\tau_{\textbf{a}+\textbf{b}}^{\gamma_0},
$$
whenever $[\tau_\textbf{a}^{\gamma_0}, \tau_\textbf{b}^{\gamma_0}]=0$. Thus, $\lambda(\tau_\textbf{a}^{\gamma_0})=1$ is a consistent context-independent value assignment. Hence there is no state-independent contextuality with HW operators for odd $d$, as claimed. 

HW-SIC thus points to a difference between odd and even $d$. As will become clear in the subsequent sections, the presence/ absence of state-independent contextuality wrt. Heisenberg-Weyl operators is linked to almost every subject in this paper.

\subsection{Summary of results}
\label{sum}

\begin{figure}
\begin{center}
\includegraphics[width=5.5cm]{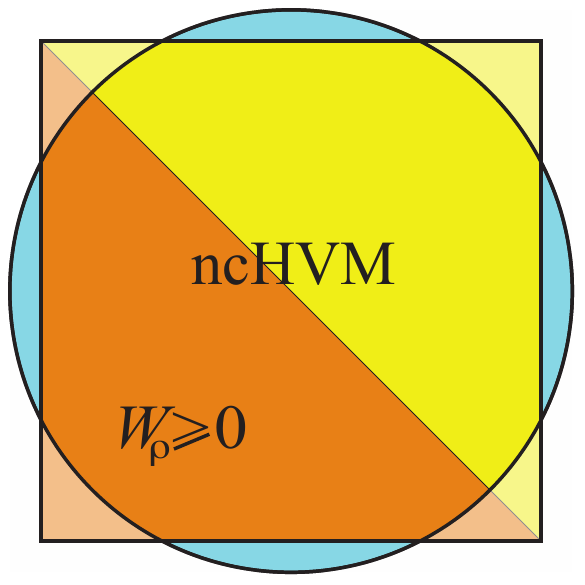}
\caption{\label{F1}Schematic view of the $n$-qubit state space. The disc represents the space of proper quantum states, the square the states describable by a non-contextual hidden variable model, and the lower triangle the states with non-negative Wigner function.}
\end{center}
\end{figure}

Our main results are the following:
\begin{enumerate}
\item{[Computational universality] For all $n$-qubit QCSI satisfying the condition (P1), contextuality of the magic states is necessary for quantum computational universality (Theorem~\ref{UnivCon}).}
\item{[Quantum speedup] For $n$-qubit QCSI schemes which satisfy the condition (P1) and for which {the value assignments of the ncHVM can be efficiently evaluated,} contextuality is necessary for speedup (Theorem~\ref{EffSimGen}).}
\item{[Existence] There is at least one family of QCSI schemes and matching Wigner function which satisfies the conditions (P1)-(P2); See Section~\ref{qubitQCSI}.}
\end{enumerate}
Points 1 and 2 also hold if ``ncHVM'' is replaced with ``non-negative Wigner function''. However, this notion of classicality is more restricted; See Fig.~\ref{F1}. In contrast to the qudit case, the existence of an ncHVM does not imply non-negativity of the Wigner function. Therefore, in the present setting, non-negativity of the Wigner function is an unnecessarily restrictive notion of classicality.

There are additional points of technical interest.
\begin{enumerate}
\item[4.]{[Preservation of classicality] If the input quantum state $\rho_\text{in}$ of an $n$-qubit QCSI satisfying condition (P1) can be described by an ncHVM, then the state of the quantum register at all later times  can be described by an ncHVM (Theorem~\ref{Res1}).} 
\item[5.]{Non-negativity of Wigner functions is in general {\em{not}} preserved under free unitary gates for QCSI schemes satisfying (P1); See Section~\ref{PP}. This does not affect efficient classical simulability (Theorem~\ref{Simul}).}
\item[6.]{For any given $n$-qubit QCSI, the state space ${\cal{S}}$ of the corresponding ncHVM consists of multiple copies of the phase space $V=\mathbb{Z}_2^n \times \mathbb{Z}_2^n$ on which the Wigner function is defined. If an $n$-qubit state is non-contextual, its representation by a probability distribution over ${\cal{S}}$ is typically not unique.}
\end{enumerate}
Comparing points 4 and 5, we find that existence of an ncHVM for a given state $\rho$ and a non-negative of Wigner function $W_\rho$ are no longer the same, as they were in the qudit case \cite{Howard}, \cite{Nicol}. Contextuality still implies Wigner negativity (Theorem~\ref{pwnc}), but not the other way around.
\smallskip

{\em{Remark:}} In our results on efficient simulation by sampling (Theorems~\ref{Simul} and \ref{EffSimGen}), we assume the sampling sources as given, and only count the operational cost of processing the samples in the simulation. This assumption holds, for example, when each magic state injected to the computation has support only on a bounded number of qubits \cite{NegWi},\cite{ReWi}. However, there is strong indication that probability distributions exist which can be efficiently prepared by quantum means but are hard to sample from classically \cite{HHT} - \cite{MH}. In view of those, Theorems~\ref{Simul} and \ref{EffSimGen} specify the computational cost of classical simulation relative to a sampling source, similar to the complexity of an algorithm relative to an oracle.

\subsection{Outline}

This paper is structured as follows. In Sections~\ref{Set} - \ref{Context} we analyze the general structure of QCSI schemes defined by the conditions (P1) - (P2), and in Section~\ref{qubitQCSI} we explicitly construct a QCSI scheme on qubits for which contextuality in the magic states is a necessary quantum mechanical resource.  Regarding the former part, in Section~\ref{Set}, we work out the implications of the conditions (P1) - (P2)  for QCSI schemes. We give a prescription for how to construct QCSI schemes starting from the phase convention $\gamma$ for the Heisenberg-Weyl operators. Section \ref{EffSim} discusses the role of Wigner functions for QCSI. In particular, we present an efficient classical simulation of QCSI for magic states with non-negative Wigner function (Algorithm 1).  Section~\ref{Context} is on the role of contextuality. We show that state-independent contextuality is absent from all QCSI schemes satisfying the conditions (P1)-(P2), clarify the relation between Wigner function negativity and state-dependent contextuality, and establish the latter as a necessary resource for QCSI with magic states. Finally, we describe an efficient classical simulation algorithm for QCSI for magic states with a non-contextual HVM  (Algorithm 2). It contains Algorithm 1 as a special case. We conclude in Section~\ref{Concl}.\medskip

We also refer to a companion paper \cite{BBW} of this article which focuses solely on the role of contextuality in QCSI. Wigner functions---and all the conceptual puzzles they give rise to in the multi-qubit setting---are bypassed. The flip side of this approach is that Wigner functions can no longer be used to characterize the near-classical ``free'' sector of operations in QCSI. Instead, the free sector is specified by  the set of available measurements, and the condition (P1) is replaced by the requirement that the available measurements do not exhibit state-independent contextuality. Ref.\ \cite{BBW} provides a shorter approach for those readers whose main interest is in contextuality.

\section{Computational setting and consistency conditions}\label{Set}

In this section we describe how qubit QCSI schemes are constructed starting from the conditions (P1), (P2). We demonstrate that, requiring (P1) and (P2), the choice of Wigner function completely determines the free sector of the corresponding QCSI scheme. Namely, (P1) defines the set ${\cal{O}}$ of observables that can be measured for free. Then, the free unitaries are the maximal set of Clifford unitaries that map the set ${\cal{O}}$ to itself under conjugation. (P2) is merely used to select the ``true'' $n$-qubit schemes. This construction is explained in Sections~\ref{Constr1}-\ref{PS}.

In Section~\ref{SetRev}, we briefly review the model of QCSI. In Section~\ref{OR}, we discuss the general concept of an operational restriction, how it overcomes the phenomenon of state-independent contextuality, and why that is necessary for establishing contextuality of the magic states as a resource for QCSI. 

\subsection{The computational setting}\label{SetRev}

Every QCSI scheme consists of four constituents, namely (i) a set $\Omega$ of states that can be prepared within the scheme (the ``free'' states), (ii) the set ${\cal{O}}$ of observables which can be directly measured, and which in the present discussion always consists solely of Pauli operators, (iii) a group $G$ of unitary gates (the ``free gates''), typically taken as the Clifford group or a subgroup thereof, and (iv) the set ${\cal{M}}$ of magic states which render the scheme computationally universal. A general QCSI scheme is thus characterized by the quadruple $({\cal{O}}, G, \Omega,{\cal{M}})$. 

The first three of these four constituents are considered ``free''. The justification for this terminology is that quantum computations built solely from the free operations cannot have a quantum speedup. This near-classicality of the free operations is made precise by an efficient classical simulation algorithm (see Section~\ref{EffSim}). It states that if the Wigner function of the initial quantum state $\rho_\text{in}$  can be efficiently sampled from then so can the outcome distribution resulting from evolving $\rho_\text{in}$ under the free unitary gates and measurements. This simulation result is the very justification for invoking a Wigner function in the description of QCSI.

\subsection{Operational restrictions}\label{OR}

When transitioning from local systems of odd prime Hilbert space dimension (qudits) to local systems of Hilbert space dimension 2 (qubits), one encounters a new phenomenon: state-independent contextuality among Pauli-observables \cite{Merm}, \cite{Entropy}. It is incompatible with viewing contextuality as a resource injected into the computation along with the magic states.

The reasons are two-fold. First, within the framework of QCSI, Pauli-mea\-sure\-ments are supposed to be free, and if contextuality is already present in those operations, how can it be a resource? Perhaps even worse, for systems of two or more qubits, a contextuality witness can be constructed that classifies {\em{all}} quantum states of $n\geq 2$ qubits as contextual \cite{Howard}, including the completely mixed state. Again, how can contextuality be a resource if it is generic?
 
In this paper, the strategy for coping with state-in\-de\-penendent contextuality is to place operational restrictions on the Pauli observables that can be measured in a QCSI scheme. The very concept of QCSI already invokes the notion of an operational restriction, since the operations in QCSI are non-universal by design. Here, additional constraints are placed by the condition (P1). The rebit case \cite{ReWi} shall serve as a model scenario for the concept of operational restrictions, and we briefly review it for illustration.

\begin{figure}
\begin{center}
\includegraphics[width=3.5cm]{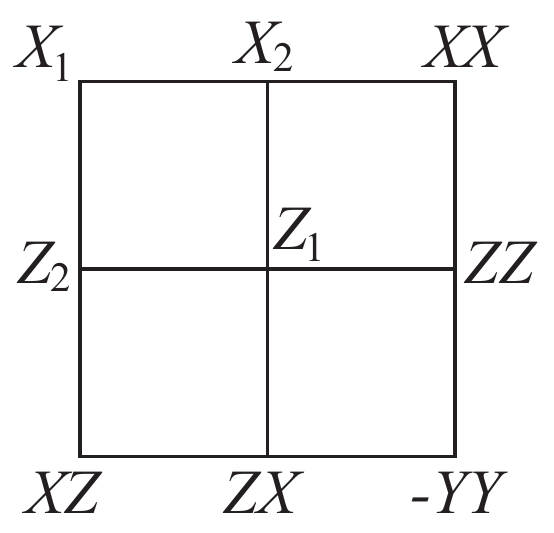}
\caption{\label{MermSq}Peres-Mermin square. For the restriction to CSS-ness preserving operations, the six observables in the top two rows are in the set ${\cal{O}}$ while the observables in the bottom row are in $M$ but not in ${\cal{O}}$. In rebit QCSI they can be measured individually but not jointly. The figure is adapted from \cite{Merm}.}
\end{center}
\end{figure}

The Peres-Mermin square \cite{Merm}, \cite{AP}, \cite{Merm3} embeds into real quantum mechanics (see Fig.~\ref{MermSq}), and confining to rebit quantum states does therefore not remove state-independent all by itself. Rather, the following operational restriction is put in place. The directly measurable observables are restricted from the set of real Pauli operators to tensor products of Pauli operators $Z_i$ only or $X_i$ only. Accordingly, the free unitaries are restricted from all real Clifford gates to those which preserve the set of Calderbank-Shor-Steane (CSS) stabilizer states \cite{CSS1},\cite{CSS2}.

Let us analyze the free measurements, for the case of $n=2$ rebits, with the CSS restriction. The set ${\cal{O}}$ of directly measurable observables is
$$
{\cal{O}} = \{I,Z_1, Z_2, Z_1Z_2, X_1, X_2, X_1X_2\}\times \{\pm 1\}
$$
By directly measuring observables from the above set, measurement outcomes of further Pauli observables can be inferred. For example, $X_1Z_2\not \in {\cal{O}}$. Yet, a value for $X_1Z_2$ can be inferred by measuring the commuting observables $X_1$ and $Z_2$ separately, and multiplying the outcomes. Applying this construction to all possible pairs of commuting observables in ${\cal{O}}$, we find the set $M$ of observables whose value can be inferred, namely
$$
M = {\cal{O}} \cup \{X_1Z_2,Z_1X_2,Y_1Y_2\} \times \{\pm 1\}.
$$ 
$M$ is thus the set of all real and Hermitian two-qubit Pauli operators. By measurement of observables in the smaller set ${\cal{O}}$ it is thus possible to fully reconstruct all two-rebit density operators.

The next question of interest is which Pauli operators can be measured {\em{jointly}}. For example, while both the observables $X_1Z_2$ and $Z_1X_2$ are in $M$ and even though they commute, in rebit QCSI they cannot have their values inferred simultaneously. Inferring the value of $X_1Z_2$ necessitates the physical measurement of the observables $X_1$ and $Z_2$, and inferring the value for $Z_1X_2$ requires the physical measurement of $Z_1$ and $X_2$. However, the four observables $X_1$, $Z_2$, $Z_1$ and $X_2$ do not all  commute. The measurement of $Z_1$ and $X_2$ to infer the outcome of $Z_1X_2$ wipes out the value of $X_1Z_2$, and vice versa.

The fact that the observables $X_1Z_2$ and $Z_1X_2$ cannot have their values inferred simultaneously is critical for state-independent contextuality. Namely, the consistency constraint among measurement outcomes for observables in the bottom row of the Mermin Peres square can no longer be experimentally checked, and is thus effectively removed from the square. As a consequence, the remaining available measurements can be described by a non-contextual hidden variable model (HVM). For example, the value assignment $\lambda=1$ for all observables in the Peres-Mermin square becomes consistent. In this way, by imposing an operational restriction, state-independent contextuality disappears from QCSI.

This concludes the review of the rebit case. In the subsequent sections we generalize the notions introduced above and apply them to a wider range of  settings. As a final remark, earlier in this section we stated that the operational restrictions must obey certain consistency conditions. The above discussion points to two of them: To give rise to a tomographically complete scheme of QCSI on qubits, the set ${\cal{O}}$ of directly measurable observables must be large enough for the derived set $M$ to comprise all Pauli operators. At the same time, ${\cal{O}}$ must be small enough to dispense with state-independent contextuality.

\subsection{Consistency conditions on $G$ and $\Omega$}\label{Constr1}

We now begin to describe the consistency conditions which must hold between the group $G$ of free unitary gates in QCSI, the set ${\cal{O}}$ of directly measurable observables, and the set $\Omega$ of free states. We require that these constituents of QCSI satisfy two constraints, namely
\begin{equation}\label{OG}
g^\dagger O g \in {\cal{O}},\;\;\forall O \in {\cal{O}},\, \forall g \in G,
\end{equation}
and
\begin{equation}\label{Om1}
 g|\psi\rangle \in \Omega,\; \forall |\psi\rangle \in \Omega,\, \forall g \in G.
\end{equation}
Regarding Eq.~(\ref{OG}), if $O$ can be measured, so can $g^\dagger O g$, namely by first applying $g$, then measuring $O$ and then applying $g^\dagger$. Likewise, if $|\psi\rangle$ can be prepared, so can $g|\psi\rangle$

We regard  the set ${\cal{O}}$ of directly measurable observables as primary among the constituents of the free sector of QCSI, and define the group $G$ of free gates and the set $\Omega$ of free states in reference to it. Namely,  $G$ is the largest subgroup of the $n$-qubit Clifford group $Cl_n$ that satisfies the property Eq.~(\ref{OG}),
\begin{equation}\label{G}
G:=\{g \in Cl_n|\,g^\dagger O g \in {\cal{O}},\,\forall O \in {\cal{O}}\}.
\end{equation}
The free states are those that can be prepared by measurement of observables in ${\cal{O}}$. All other states are considered resources, and must be provided externally if needed. That is, $|\psi\rangle \in \Omega$ if and only if there exists an ordered set ${\cal{O}}_{|\psi\rangle} \subset {\cal{O}}$ such that
\begin{equation}\label{Om}
|\psi\rangle \langle \psi| \sim \left(\prod_{O \in {\cal{O}}_{|\psi\rangle}} \left[ \frac{I\pm O}{2}\right]\right) (I/2^n).
\end{equation}
The projectors on the lhs. of Eq.~(\ref{Om}) do not necessarily commute. Their temporal order is specified by the ordering in ${\cal{O}}_{|\psi\rangle}$. The angular brackets denote superoperators. With Eq.~(\ref{G}), ${\cal{P}}_n \subset G$ always holds. Therefore, a totally depolarizing  twirl may be implemented, producing $I/2^n$ from any $n$-qubit state.

The free sector of a QCSI scheme is thus fully specified via Eqs.~(\ref{G}) and (\ref{Om}) by the set ${\cal{O}}$ of directly measurable observables. In Section~\ref{ConsCond} we turn to the question of how ${\cal{O}}$ itself is constructed.

\subsection{Wigner functions}\label{WiFu}
 
A Wigner function is a means of description of QCSIs. The reason for invoking Wigner functions is to characterize the near-classicality of the sector of free operations in QCSI. This proceeds by way of the efficient classical simulation algorithm described in Section~\ref{SimulAlg}.

The Wigner functions considered here are defined on a phase space $V := \mathbb{Z}_2^n \times \mathbb{Z}_2^n$, starting from the Heisenberg-Weyl operators 
\begin{equation}\label{Tdef}
T^\gamma_\textbf{a}=i^{\gamma(\textbf{a})}Z(\textbf{a}_Z)X(\textbf{a}_X).
\end{equation}
Therein, $Z(\textbf{a}_Z):=\bigotimes_{i=1}^n {Z_i}^{\textbf{a}_{Z,i}}$, $X(\textbf{a}_X):=\bigotimes_{i=1}^n {X_i}^{\textbf{a}_{X,i}}$.

The possible phase conventions $\gamma: V \longrightarrow \mathbb{Z}_4$ are constrained only by the requirement that all $T^\gamma_\textbf{a}$, $\textbf{a} \in V$, are Hermitian. As we show later, the QCSI schemes considered here and the Wigner functions describing them are both fully specified by $\gamma$.

We consider Wigner functions of the form $W^\gamma_\rho(\textbf{u}) = 1/2^n\, \text{Tr} (A_\textbf{u} \rho)$, for all $\textbf{u} \in V = \mathbb{Z}_2^{2n}$, where $A_\textbf{u} = T^\gamma_\textbf{u} A_\textbf{0} (T^\gamma_\textbf{u})^\dagger$,
\begin{equation}\label{Adef}
A^\gamma_\textbf{0} =\frac{1}{2^n} \sum_{\textbf{a} \in V}T^\gamma_\textbf{a}.
\end{equation}

This definition satisfies the minimal conditions required of a Wigner function \cite{Gibbons}, namely that (i) $W^\gamma$ is a quasi-probability distribution defined on a state space $V=\mathbb{Z}_2^{2n}$, (ii) $W^\gamma$ transforms covariantly under the Pauli group, $W^\gamma_{T_\textbf{a}\rho T_\textbf{a}^\dagger}(\textbf{u}) = W_\rho(\textbf{u}+\textbf{a})$, for all $\textbf{u},\textbf{a} \in V$, and (iii) there is a suitable notion of marginals. 

{\em{Remark:}} To simplify the notation, we subsequently omit the superscript $\gamma$ in the Wigner functions, unless the precise choice of $\gamma$ matters.\medskip

All previous works on the role of positive Wigner functions for QCSI---\cite{Galv}, \cite{Galv2}, \cite{NegWi}, \cite{ReWi}---are based on a particular family of Wigner functions for finite-dimensional state spaces introduced by Gibbons {\em{et al.}} \cite{Gibbons}. This is, indirectly, also the case for the present Wigner function, and we therefore briefly describe its genealogy. Gibbons {\em{et al.}} introduced a family of Wigner functions for finite-dimensional state spaces based on the concepts of mutually unbiassed bases and lines in phase space. Among this family, for the special case of odd local dimension, Gross \cite{Gross} identified a Wigner function which is the most sensible finite-dimensional analogue of the infinite-dimensional case \cite{Wigner}. This Wigner function was written in the form of Eqs.~(\ref{Tdef}),(\ref{Adef}) in \cite{NegWi}, with a special phase convention $\gamma$. For local Hilbert space dimension 2, this special function $\gamma$ does not exist, and in the present approach $\gamma$ is left as a parameter to vary. The freedom of choosing the function $\gamma$ replaces the freedom of choosing quantum nets in \cite{Gibbons}.\medskip

In addition to the above Properties (i) - (iii), the Wigner functions defined in Eqs.~(\ref{Tdef}), (\ref{Adef}) have two further relevant properties. First, for any pair $\rho$ and $\sigma$ of operators acting on the Hilbert space $\mathbb{C}^{2^n}$, it holds that
\begin{equation}\label{Trip}
\text{Tr}(\rho \sigma) = 2^n \sum_{\textbf{u} \in V} W_\rho(\textbf{u}) W_\sigma(\textbf{u}).
\end{equation}
Second, for any admissible function $\gamma$, we have the following relation between a quantum state $\rho$ and its Wigner functions $W_\rho$,
$$
\rho = \sum_{\textbf{u} \in V}W_\rho(\textbf{u}) A_\textbf{u}.
$$

\subsection{{The condition (P1) defines ${\cal{O}}$}}\label{ConsCond}

The set ${\cal{O}}$ of directly measurable observables is, in the present setting, always a set of Hermitian Pauli operators,
$$
{\cal{O}}=\{\pm T_\textbf{a} , \textbf{a} \in V_{\cal{O}}\},
$$
where $V_{\cal{O}}$ is a subset of $V=(\mathbb{Z}_2)^{2n}$.

In Section~\ref{Constr1} we described how to construct the set $\Omega$ of free states given the set ${\cal{O}}$ of directly measurable observables. But how is the set ${\cal{O}}$ itself constructed? 
To answer this question, we return to the function $\gamma$ in Eq.~(\ref{Tdef}) from which everything follows in the present setting. 
The function $\gamma: V \longrightarrow \mathbb{Z}_4$ specifies  a function $\beta: V \times V \longrightarrow \mathbb{Z}_4$ defined via
\begin{equation}\label{3T}
T_{\textbf{a}+\textbf{b}}=i^{\beta(\textbf{a},\textbf{b})}T_\textbf{a}T_\textbf{b}.
\end{equation}
{The function $\beta$  is related to the function $\gamma$ introduced in Eq.~(\ref{Tdef}) via
\begin{equation}\label{betagamma}
\beta(\textbf{a},\textbf{b}) = -\gamma(\textbf{a})-\gamma(\textbf{b}) +\gamma(\textbf{a}+\textbf{b}) + 2\textbf{a}_X\textbf{b}_Z \mod4.
\end{equation}
Thus, $\beta$ is fully determined by $\gamma$. The converse is not true. Every valid function $\beta$ (i.e., one that derives from a function $\gamma$) determines $\gamma$ only up to a translation in phase space.}

The function $\beta$ constrains the Pauli operators that can possibly be contained in the set ${\cal{O}}$. Namely, we have the following Lemma.
\begin{Lemma}\label{InO}
For any $\textbf{a} \in V$, the measurement of an observable $\pm T_\textbf{a}$ does not introduce negativity into the Wigner function if and only if 
\begin{equation}\label{C1}
\beta(\textbf{a},\textbf{b})= 0,\; \forall \textbf{b} \in V|\, [\textbf{a},\textbf{b}]=0.
\end{equation}
\end{Lemma}
In Eq.~(\ref{C1}), $[\cdot,\cdot]$ is the symplectic bilinear form defined by $[\textbf{a},\textbf{b}]:=\textbf{a}_X\cdot \textbf{b}_Z+ \textbf{a}_Z\cdot\textbf{b}_X \mod 2$, for all $\textbf{a},\textbf{b} \in V$.\smallskip
 
{\em{Proof of Lemma~\ref{InO}.}}  ``Only if'': Assume that the condition Eq.~(\ref{C1}) does not hold, i.e., there exists a Pauli operator $T_\textbf{b}$ such that $[\textbf{a},\textbf{b}]=0$ and  $\beta(\textbf{a},\textbf{b})\neq 0=2$ (Hermiticity). 

Further assume that the system is in the mixed state $(I - T_\textbf{b})/2^n$, which has  non-negative $W$, and that $T_\textbf{a}$ is measured. W.l.o.g. assume that the outcome is -1. The resulting state is $\rho = (I-T_\textbf{a}-T_\textbf{b}+T_\textbf{a}T_\textbf{b})/2^n=(I-T_\textbf{a}-T_\textbf{b}-T_{\textbf{a}+\textbf{b}})/2^n$. Thus, $W_\rho(\textbf{0}) =-2/4^n <0$.  Thus, if $\beta(\textbf{a},\textbf{b})\neq 0$ for some $\textbf{b} \in V$, the measurement of $T_\textbf{a}$ can introduce negativity into Wigner functions, hence $\pm T_\textbf{a} \not \in {\cal{O}}$. Negation of this statement proofs the result. \smallskip

``If'': We assume that the Wigner function $W_\rho$ of the state $\rho$ before the measurement is non-negative, 
$$W_\rho(\textbf{u}) \geq 0,\; \forall \textbf{u} \in V,$$
and that the measured observable $T_\textbf{a}$ is such that $\beta(\textbf{a},\textbf{b}) = 0$, for all $\textbf{b} \in V$.
The state $\rho'$ after the measurement of the observable $T_\textbf{a}$ with outcome $s \in \{0,1\}$ is $\rho' \sim \frac{I + (-1)^s T_\textbf{a}}{2} \rho  \frac{I + (-1)^s T_\textbf{a}^\dagger}{2}$, and the value of the corresponding Wigner function at the phase space point $\textbf{u} \in V$ is
\begin{equation}\label{pW}
p_\textbf{a}(s) W_{\rho'}(\textbf{u}) = \frac{1}{2^n} \text{Tr} \left(\frac{I + (-1)^s T_\textbf{a}^\dagger}{2} A_\textbf{u}  \frac{I + (-1)^s T_\textbf{a}}{2}\, \rho \right).
\end{equation}
Therein, $p_\textbf{a}(s)$ is the probability of obtaining the outcome $s$ in the measurement of $T_\textbf{a}$.
Now,
$$
\begin{array}{l}
\displaystyle{\frac{I + (-1)^s T_\textbf{a}^\dagger}{2} A_\textbf{u}  \frac{I + (-1)^s T_\textbf{a}}{2}} =\\
\; = \displaystyle{\frac{I + (-1)^s T_\textbf{a}^\dagger}{2} \left(\frac{1}{2^n} \sum_{\textbf{b}\in V}  (-1)^{[\textbf{u},\textbf{b}]} T_\textbf{b} \right) \frac{I + (-1)^s T_\textbf{a}}{2}}\\
\; =  \displaystyle{\frac{I + (-1)^s T_\textbf{a}}{2} \left(\frac{1}{2^n} \sum_{\textbf{b}\in V| [\textbf{b},\textbf{a}]=0}  (-1)^{[\textbf{u},\textbf{b}]} T_\textbf{b} \right)}\\
\; =  \displaystyle{\frac{1}{2^{n+1}} \sum_{\textbf{b}\in V| [\textbf{b},\textbf{a}]=0}  (-1)^{[\textbf{u},\textbf{b}]} \big(T_\textbf{b}+(-1)^s T_\textbf{a}T_\textbf{b}\big)}\\
\; =  \displaystyle{\frac{1}{2^{n+1}} \sum_{\textbf{b}\in V| [\textbf{b},\textbf{a}]=0}  (-1)^{[\textbf{u},\textbf{b}]} \big(T_\textbf{b}+(-1)^s T_{\textbf{a}+\textbf{b}}\big)}\\
\; =  \displaystyle{\frac{1}{2^{n+1}} \sum_{\textbf{b}\in V| [\textbf{b},\textbf{a}]=0}  (-1)^{[\textbf{u},\textbf{b}]}  \left(1+(-1)^{s+[\textbf{a},\textbf{u}]} \right) T_\textbf{b} }\\
\; =  \displaystyle{\frac{\delta_{s,[\textbf{a},\textbf{u}]}}{2^{n}} \sum_{\textbf{b}\in V| [\textbf{b},\textbf{a}]=0}  (-1)^{[\textbf{u},\textbf{b}]}  T_\textbf{b} }\\
\; = \displaystyle{\frac{\delta_{s,[\textbf{a},\textbf{u}]}}{2} \left(A_\textbf{u} + A_{\textbf{u} +\textbf{a}} \right)}.
\end{array}
$$
Above, we have used the assumption that $\beta(\textbf{a},\textbf{b}) =0 $ for all $\textbf{b} \in V$ when transitioning from line 4 to line 5. Applying the result to Eq.~(\ref{pW}) we find that
\begin{equation}\label{WigUp}
p_\textbf{a}(s) W_{\rho'}(\textbf{u}) = \frac{\delta_{s,[\textbf{a},\textbf{u}]}}{2} \big(W_\rho(\textbf{u}) + W_\rho(\textbf{u} +\textbf{a}) \big).
\end{equation}
By assumption, the r.h.s. is always non-negative. For the outcome $s$ to possibly occur, it is required that $p_\textbf{a}(s) > 0$. Hence, $W_{\rho'}(\textbf{u}) \geq 0$, for all $\textbf{u} \in V$. $\Box$\medskip

{We define the set ${\cal{O}}$ of directly measurable observables to be the largest possible set of Pauli observables allowed by Lemma~\ref{InO},
\begin{equation}\label{Omax}
{\cal{O}}: = \{ \pm T_\textbf{a},\, \textbf{a} \in V|\, \beta(\textbf{a},\textbf{u})=0,\, \forall \textbf{u} \in V_\textbf{a}\},
\end{equation}
with $V_\textbf{a}:=\{\textbf{u}\in V| [\textbf{a},\textbf{u}]=0\}$.}\medskip

{\em{Example:}} To illustrate the usefulness of Lemma~\ref{InO}, consider the following choice for $\gamma$. For brevity, we restrict to two rebits. $W^\gamma$ is specified by$$
\begin{array}{rcl}
A^\gamma_\textbf{0} &=& \frac{1}{4} \big(I - Z_1 + Z_2 + Z_1Z_2 - X_1+ X_2 + X_1X_2 + \\
&& + X_1Z_2 + Z_1X_2 - Y_1Y_2 \big).
\end{array}
$$
Arranging all observables in $A_0$ apart from the identity in the Peres-Mermin square,
$$
\includegraphics[width=3.9cm]{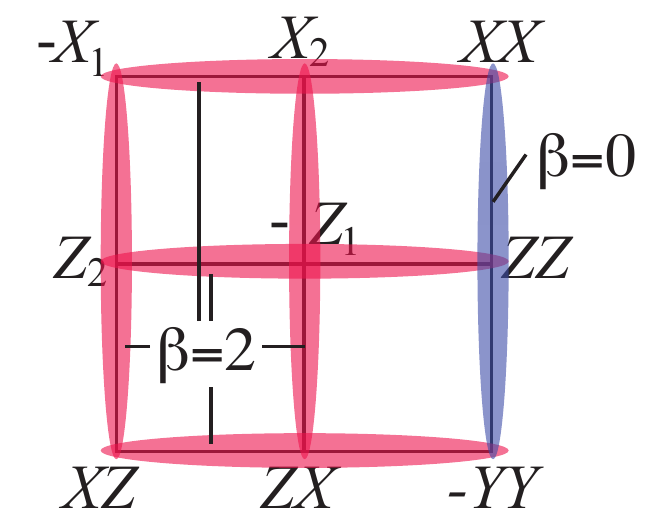}
$$
it is evident that every observable $T_\textbf{a}$ is part of at least one commuting triple with $\beta \neq 0$. Hence, apart from the identity, no observable is in ${\cal{O}}$, i.e., ${\cal{O}}=\{ I\}$.  The corresponding QCSI scheme is thus the exact opposite of tomographically complete: Nothing can be measured at all!  We find that not for every function $\gamma$ the Wigner function $W^\gamma$ can be paired with a matching QCSI scheme.\medskip 

For further illustration of Lemma~\ref{InO}, we have the following implication.
\begin{Lemma}\label{NotAll}
Consider a Wigner function as defined in Eqs.~(\ref{Adef}), (\ref{Tdef}), for $n\geq 2$ qubits. Then, there always exists a Pauli observable whose measurement does not preserve positivity. 
\end{Lemma}
Thus, no QCSI scheme in which all Pauli observables are directly measurable can satisfy the property (P1). The original QCSI scheme \cite{BK} on qubits is one of those schemes, and it is therefore out of scope of the present analysis.

{\em{Proof of Lemma~\ref{NotAll}.}} Consider the Peres-Mermin square as displayed in Fig.~\ref{MermSq}. Irrespective of the sign conventions of the Pauli observables contained in it, there is always at least one context with $\beta=2$. Thus, by Lemma~\ref{InO}, for $n\geq 2$ the measurement of at least three Pauli observables introduces negativity into the Wigner function. $\Box$\medskip

Furthermore, regarding the free states, an immediate consequence of Property (P1) is that all free states $|\Psi\rangle \in \Omega$ are non-negatively represented by $W$. All free states $|\Psi\rangle \in \Omega$ can be created from the completely mixed state $I/2^n$,  by measurement of observables in ${\cal{O}}$, and $W_{I/2^n} \geq 0$ for any $\gamma$. Then, with Property (P1), $W_{|\Psi\rangle \langle \Psi|}\geq 0$, for all $|\Psi\rangle \in \Omega$. By Eq.~(\ref{Om1}), free states remain non-negatively represented upon action of free unitary gates $g\in G$.

\subsection{Inferability and tomographic completeness}

{We know from Lemma~\ref{NotAll} that for $n\geq 2$ qubits, the set ${\cal{O}}$ of directly measurable observables is always strictly smaller than the set of all Pauli observables. How is that not in conflict with tomographic completeness (P2)? }  

{The reason---as was already mentioned in Section~\ref{OR}---is that some Pauli observables, while not directly measurable, can nonetheless have their value inferred by measurement. For example, consider three Pauli observables $T_a,T_b\in {\cal{O}}$, $T_c\not\in {\cal{O}}$, such that $[a,b]=0$ and $T_c = T_aT_b$. Then $T_c$ can have its value inferred by measuring $T_a$ and $T_b$, and multiplying the outcomes.}

{For suitable sets ${\cal{O}}$, all Pauli observables can have their value inferred, even if not directly measured. This suffices to satisfy (P2).}

\begin{Def}\label{Mdef}
$M = \{\pm T_\textbf{a}|\, \textbf{a} \in V_M \}$ is the set of Pauli observables whose value can be inferred from a {\em{single copy}} of the given quantum state, by measurement of other Pauli observables from the set ${\cal{O}}$ and classical post-processing.
\end{Def}
The set $M$ is typically larger than the set ${\cal{O}}$ of observables which can be directly measured. This was illustrated by an example in Section~\ref{OR}, namely ${\cal{O}}=\{I,X_1,Z_2\}$, $M={\cal{O}} \cup \{X_1Z_2\}$. In terms of $V_M$, the condition (P2) of tomographic completeness reads
\begin{equation}\label{C2}
V_M = V.
\end{equation}
We now provide a general characterization of the set $M$ generated by the set ${\cal{O}}$.
\begin{Lemma}\label{VMchar}
For any $\gamma$, the set $V_M$ has the properties that (i) $V_{\cal{O}} \subseteq V_M$, and (ii) for any $\textbf{a} \in V_{\cal{O}}$, $\textbf{b} \in V$ with $[\textbf{a},\textbf{b}]=0$, it holds that $\textbf{a} + \textbf{b} \in V_M$ if and only if $\textbf{b}\in V_M$. 
\end{Lemma}

{\em{Proof of Lemma~\ref{VMchar}.}} Property (i) merely states that what can be directly measured can have its value inferred. Regarding (ii), the observable $T_{\textbf{a}+\textbf{b}}$ has its value inferred as follows. First, $T_\textbf{a} \in {\cal{O}}$ is measured directly. Then, the procedure for inferring the value of $T_\textbf{b}$ is applied. Since $T_\textbf{a}$ commutes with $T_\textbf{b}$, the former measurement doesn't interfere with the latter, and $\mu(T_\textbf{a}T_\textbf{b}) = \mu(T_\textbf{a})\mu(T_\textbf{b})$. Finally, with Eq.~(\ref{C1}), $\mu(T_{\textbf{a}+\textbf{b}}) = \mu(T_\textbf{a})\mu(T_\textbf{b})$. Thus, if $\textbf{b}\in V_M$ then $\textbf{a}+\textbf{b} \in V_M$. The reverse direction holds by symmetry in $\textbf{b}\longleftrightarrow \textbf{a}+\textbf{b}$. $\Box$\medskip

{\em{Example.}} Assume that  $X_1,Z_2,Y_1Y_2 \in {\cal{O}}$. The  outcome of the observable $Z_1X_2$ can then be inferred by measurement, i.e. $Z_1X_2 \in M$. 
The procedure for the measurement of the observable $Z_1X_2$, given the above set ${\cal{O}}$ of directly measurable observables, is the following. First, the observable $Y_1Y_2$ is measured, and second the commuting observables $X_1$ and $Z_2$ are measured. The measurement outcome $\mu(Z_1X_2) \in \{\pm 1\}$ then is 
$$
\mu(Z_1X_2) = \mu(Y_1Y_2)\mu(X_1)\mu(Z_2).
$$
The key point of this example is that not all pairs among the measured Pauli observables $X_1$, $Z_2$ and $Y_1Y_2$ commute; yet in the above expression for $\mu(Z_1X_2)$ we treated them as if they did. The reason that this is possible is the following. 

Since $Y_1Y_2$ does not commute with $X_1$ and $Z_2$, the measurements of $X_1$ and $Z_2$ after the measurement of $Y_1Y_2$---if taken separately---do not reveal any information about the initial state. Individually, their outcomes are completely random, whatever the state prior to the $Y_1Y_2$-measurement is. However, $X_1$ and $Z_2$ mutually commute, and hence the separate measurement of $X_1$ and $Z_2$ implies a valid measurement outcome for the correlated observable $X_1Z_2$, namely $\mu(X_1Z_2)=\mu(X_1)\mu(Z_2)$. Furthermore, since $X_1Z_2$ does commute with $Y_1Y_2$, $\mu(X_1)\mu(Z_2)$ represents the outcome of a $X_1Z_2$-measurement on the initial state, and  $\mu(Z_1X_2)=\mu(Y_1Y_2)\mu(X_1Z_2)=\mu(Y_1Y_2)\mu(X_1)\mu(Z_2)$, as claimed.

Let us now verify that $Z_1X_2 \in M$ follows from the properties established in Lemma~\ref{VMchar}. First, with Property (i) of Lemma~\ref{VMchar}, $X_1 \in {\cal{O}}$ implies $X_1 \in M$. Then, using Property (ii) with $X_1 \in M$, $Z_2\in {\cal{O}}$, it follows that $X_1Z_2 \in M$. Finally, again with Property (ii), since $Y_1Y_2 \in {\cal{O}}$ and $X_1Z_2 \in M$, it follows that $Z_1X_2 \in M$.

We note that the above procedure of inferring measurement outcomes by the physical measurement of non-commuting observables is reminiscent of the syndrome measurement in subsystem codes \cite{SSC1}, \cite{SSC2}, with the Bacon-Shor code \cite{Bac}, \cite{ShorC} and topological subsystem codes \cite{Bombin} as prominent examples.

\smallskip

Back to the general scenario, an observable $T_\textbf{a}$ can have its value inferred, i.e., $T_\textbf{a} \in M$, if there exists a resolution
\begin{equation}\label{resol}
\textbf{a} = \textbf{a}_1 + \left( \textbf{a}_2 + \left( \textbf{a}_3 + ... (\textbf{a}_{N-1}+\textbf{a}_N)...\right)\right),
\end{equation}
where all $\textbf{a}_i \in V_{\cal{O}}$, and 
\begin{equation}\label{CommReq}
\left[\textbf{a}_i,\sum_{j=i+1}^N\textbf{a}_j\right]=0,\; \forall i=1,..,N-1.
\end{equation} 
The resolution Eq.~(\ref{resol}) of $\textbf{a}$ represents a measurement  
sequence for inferring the value of $T_\textbf{a}$, starting with the measurement of $\textbf{a}_1$ and ending with the measurement of $\textbf{a}_N$. The inferred value is $\lambda(T_\textbf{a}) = \prod_{i=1}^N \lambda(T_{\textbf{a}_i})$.\medskip

Finally, we introduce a generalization of the set $M$ of Pauli observables whose value can be inferred. Namely, we denote by $C$, $C \subset M$, a set of observables whose value can be inferred {\em{jointly}} in QCSI. For short, we call such a set $C$ a ``set of jointly measurable observables''.
\begin{Def}\label{DefC} A set $C$, $C \subset M$, of commuting Pauli observables is jointly measurable if  the outcomes for all observables in $C$ can be simultaneously inferred from measurement of observables in ${\cal{O}}$ on a {\em{single copy}} of the given quantum state $\rho$. 
\end{Def}
The sets $C$ of simultaneously measurable observables will become important in Section~\ref{NWNC}, where we discuss the relation between contextuality and negativity of Wigner functions. 

Some examples for possible sets $C$ are (i) $C=\{O\}$, for any $O \in M$, (ii) any commuting subset of ${\cal{O}}$, and (iii) $C=\{A,B,AB\}$, for $A \in M$, $B \in {\cal{O}}$ and $[A,B]=0$.

We have the following characterization of the sets $C$ of simultaneously measurable observables.
\begin{Lemma}\label{C+}
Consider a set $C$ of simultaneously measurable observables, and $T_\textbf{a},T_\textbf{b} \in C$. Then, $T_{\textbf{a}+\textbf{b}} = T_\textbf{a} T_\textbf{b}$.
\end{Lemma}
We postpone the proof of Lemma~\ref{C+} until Section~\ref{NoSIC}.

\subsection{Constructing QCSI schemes from $\gamma$}\label{PS}

Once the phase convention $\gamma$ for the $n$-qubit Pauli operators is given (cf. Eq.~(\ref{Tdef})), the Wigner function Eq.~(\ref{Adef}) and the free sector of the corresponding QCSI scheme satisfying (P1) are fully specified. They are obtained through the following steps:
\begin{enumerate}
\item{Construct the Wigner function $W$ via its definition Eqs.~(\ref{Tdef}), (\ref{Adef}).}
\item{Compute the function $\beta$ defined in Eq.~(\ref{3T}) from the function $\gamma$. Construct the set ${\cal{O}}$ of directly measurable observables via Eq.~(\ref{Omax}).} 
\item{Construct the group $G$ of free unitary gates via Eq.~(\ref{G}), and set $\Omega$ of free states via Eq.~(\ref{Om}).} 
\end{enumerate}
In addition, for tomographic completeness (P2) it needs to be checked that every Pauli observable $T_\textbf{a}$ has a resolution Eq.~(\ref{resol}).\medskip

To summarize, in this section we have stated minimal requirements for any QCSI scheme on qubits and its corresponding Wigner function. We have shown that once the function $\gamma$ is provided, the free sector of the corresponding QCSI scheme is fully specified. {To reflect this fact in our notation, we subsequently denote QCSI schemes by $(\gamma,{\cal{M}})$ instead of $({\cal{O}},G,\Omega,{\cal{M}})$. Implicit in this notation is that matching pairs of a QCSI scheme $(\gamma,{\cal{M}})$ and a Wigner function $W^\gamma$ satisfy (P1).}

\section{Efficient classical simulation of QCSI for non-negative states}\label{EffSim}

The purpose of this section is to clarify the role of Wigner functions for QCSI schemes satisfying the condition (P1). Wigner functions endow the free operations with a notion of classicality, based on efficient simulability by sampling. In Sections~\ref{Reform} and \ref{SimulAlg} we demonstrate the following result.

\begin{Theorem}\label{Simul}
{For any QCSI scheme $(\gamma, {\cal{M}})$,} if (i) the Wigner function $W^\gamma_{\rho_\text{in}}$ of the initial state $\rho_\text{in}$ can be efficiently sampled from, and (ii) the phase convention $\gamma(\textbf{a})$ can be efficiently evaluated for all $\textbf{a} \in V_{\cal{O}}$, then the distribution of measurement outcomes can be efficiently sampled from.
\end{Theorem}
In so far, Wigner functions do the same work as in the qudit case \cite{NegWi}. However, as we show in Section~\ref{Disc}, there also are differences. For example, the free unitary gates may introduce negativity into the Wigner function at hand, without compromising efficient classical simulability. The negativity of Wigner functions is thus no witness of quantumness.

\subsection{Reformulation of the simulation problem}\label{Reform}

For the purpose of classical simulation, we make an alternation to the present QCSI scheme, which, however, does not affect its computational power. Namely, we absorb the unitary gates in the measurements, such that only state preparations and measurements remain as free operations. Here we take the viewpoint that all there is to simulate about a quantum computation is the joint outcome distribution of measurements performed in course of the computation. If the unitaries can be removed without altering the outcome distribution, then there is certainly no loss in removing them. But there is a gain: as is made explicit in Section~\ref{Disc}, the simulation algorithm of Section~\ref{SimulAlg} can handle free unitaries $g \in G$ that introduce negativity into the Wigner function of the processed quantum state.

\begin{figure}
\includegraphics[width=8cm]{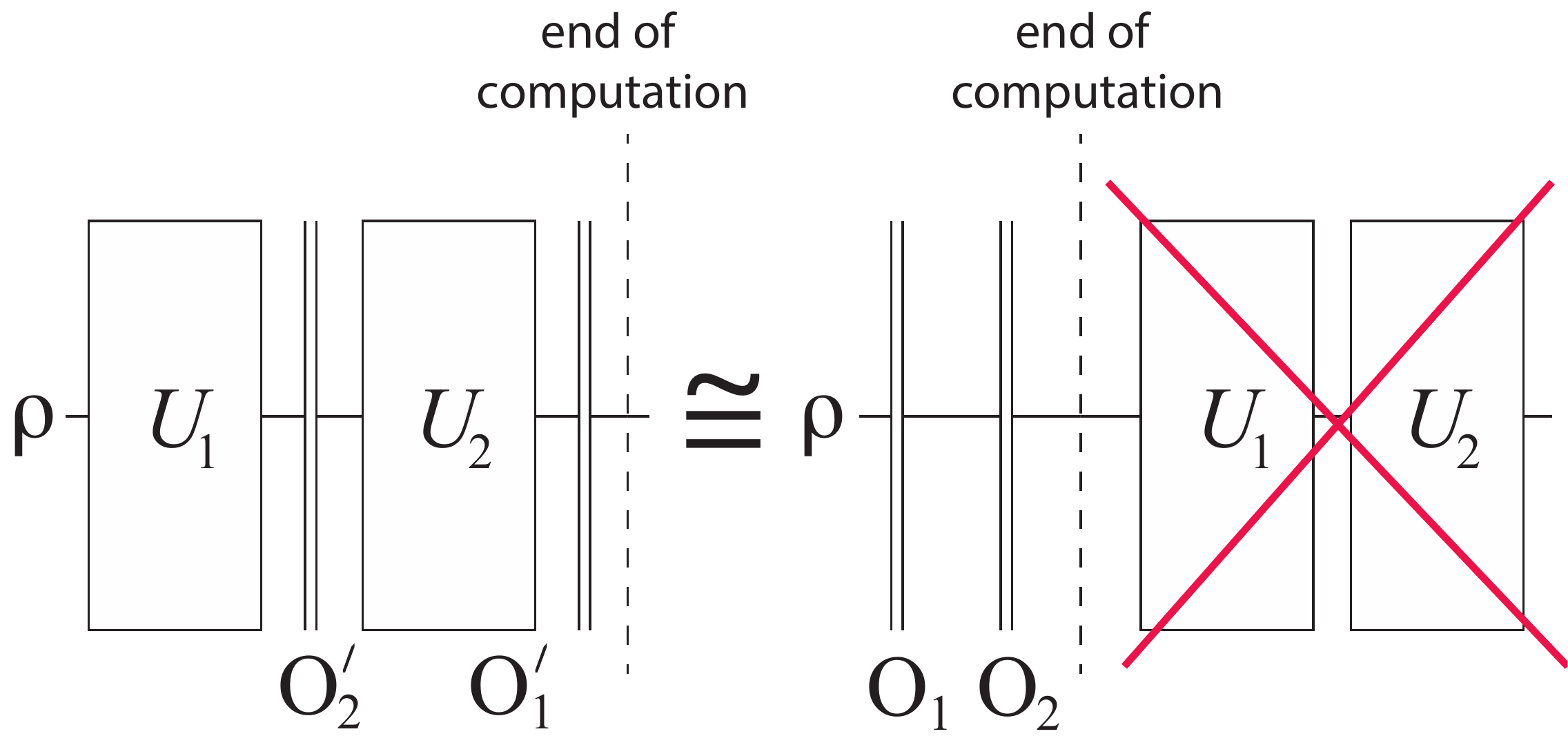}
\caption{\label{Conj} The measurements of observables $O'_i$ are propagated backwards in time to act on the initial state, by conjugation under the interspersed unitaries. Since only the measurement statistics is of interest, the trailing unitaries may  be removed from the resulting circuit.}
\end{figure}

The general procedure is outlined in Fig.~\ref{Conj}. Consider a QCSI circuit which is an alternation of unitary gates $g_i(\textbf{s}_{\prec i})$ and projective measurements represented by projectors $P'_i(\textbf{s}_{\prec i}, s_i)$,
\begin{equation}\label{circ}
{\cal{C}} = \prod_{i=1}^{t_\text{max}} P'_i(\textbf{s}_{\prec i}, s_i) g_i(\textbf{s}_{\prec i}),
\end{equation}
where the factors in the product are ordered from right to left. Therein, $\textbf{s}$ is the binary vector of all measurement outcomes, and $\textbf{s}_{\prec i}$ is $\textbf{s}$ restricted to the measurement outcomes obtained prior to the gate $g_i$. We thus allow unitary gates and measurements to depend on previously obtained measurement outcomes. Such conditioning is essential for the working of QCSI.

Now denote by $G_i(\textbf{s}_{\prec i})$ the unitaries accumulated up to step $i$,
\begin{equation}\label{AccumUni}
G_i(\textbf{s}_{\prec i}) = \prod_{j=1}^i g_i(\textbf{s}_{\prec i}).
\end{equation}
Therein, the ordering of operations is the same as in Eq.~(\ref{circ}). The circuit ${\cal{C}}$ of Eq.~(\ref{circ}) may then be rewritten as
$$
\begin{array}{rcl}
{\cal{C}} &=& \displaystyle{\prod_{i=1}^{t_\text{max}} P'_i(\textbf{s}_{\prec i}, s_i) g_i(\textbf{s}_{\prec i})}\\
&=&  \displaystyle{G_{t_\text{max}}(\textbf{s}) \prod_{i=1}^{t_\text{max}}  G_i(\textbf{s}_{\prec i})^\dagger P'_i(\textbf{s}_{\prec i}, s_i) G_i(\textbf{s}_{\prec i})}\\
&\cong&  \displaystyle{\prod_{i=1}^{t_\text{max}}  G_i(\textbf{s}_{\prec i})^\dagger P'_i(\textbf{s}_{\prec i}, s_i) G_i(\textbf{s}_{\prec i})}.
\end{array}
$$
Thus, if the measured observables in the original sequence of operations were $O'_i(\textbf{s}_{\prec i})$, the corresponding observables in the equivalent sequence are
\begin{equation}\label{ObsRel}
O_i(\textbf{s}_{\prec i}) = G_i(\textbf{s}_{\prec i})^\dagger O'_i(\textbf{s}_{\prec i}) G_i(\textbf{s}_{\prec i}).
\end{equation}
By Eq.~(\ref{OG}), if $O'_i(\textbf{s}_{\prec i}) \in {\cal{O}}$, then  $O_i(\textbf{s}_{\prec i}) \in {\cal{O}}$. Therefore, a QCSI scheme with set ${\cal{O}}$ of measurable observables and group $G$ of unitary gates is equivalent to a QCSI scheme with set ${\cal{O}}$ of measurable observables and no unitaries at all.

\subsection{Simulation algorithm}\label{SimulAlg}

The classical simulation algorithm for the setting of Theorem~\ref{Simul} is given in Table~\ref{Algo1}.

\begin{table}
\begin{tabular}{l}
\textbf{Algorithm 1}\\ \hline\hline 
\parbox{8cm}{
\begin{enumerate}
\item{Draw a sample $\textbf{v}\in V$ from $W_{\rho_\text{in}}$, and set $\textbf{v}_1:=\textbf{v}$.}
\item{For all the measurements of observables $T_{\textbf{a}_i} \in {\cal{O}}$ comprising the circuit, starting with the first,
\begin{enumerate}
\item{Output the result $s_i = [\textbf{a}_i,\textbf{v}_i]$ for the measurement of the observable $T_{\textbf{a}_i}$,}
\item{Flip a fair coin, and update the sample 
$$
\textbf{v}_i \longrightarrow \textbf{v}_{i+1} = \left\{\begin{array}{ll}  \textbf{v}_i,&\text{if ``heads''} \\ \textbf{v}_i +\textbf{a}_i,& \text{if ``tails'} \end{array} \right. ,
$$}
\end{enumerate}
until the measurement sequence is complete.}
\item{Repeat until sufficient statistics is gathered.}
\end{enumerate}} \\ \hline
\end{tabular}
\caption{\label{Algo1}Algorithm 1 for the classical simulation of $n$-qubit QCSI with non-negative Wigner function of the initial state.}
\end{table}

Any sample $\textbf{u} \in V$ from a non-negative Wigner function has a definite value assignment for all observables in the Pauli group. Namely, for any $\textbf{a} \in V$, the measurement outcome for the Pauli observable $T_\textbf{a}$ is
\begin{equation}\label{ValAss}
\lambda_\textbf{u}(\textbf{a}) = (-1)^{[\textbf{a},\textbf{u}]}.
\end{equation} 
The value assignment Eq.~(\ref{ValAss}) is a direct consequence of the update rule Eq.~(\ref{WigUp}). Namely, in the l.h.s. of Eq.~(\ref{WigUp}) assume that the probability $p_\textbf{a}(s)$ for obtaining the outcome $s\in \{0,1\}$ in a measurement of a Pauli observable $T_\textbf{a}$ on a state $\rho$ is non-zero. Then, the r.h.s. of Eq.~(\ref{WigUp}) implies that $s=[\textbf{a},\textbf{u}]$, or, equivalently, $\lambda_\textbf{u} = (-1)^{[\textbf{a},\textbf{u}]}$.

For illustration of the state update rule in the above classical simulation algorithm, consider two measurement sequences for the state of one qubit, namely (i) Repeated measurements of the Pauli observable $Z=\pm T_{\textbf{a}(Z)}$, and (ii) Alternating measurements of the Pauli observables $Z$ and $X=\pm T_{\textbf{a}(X)}$. Assume that the sample from the Wigner function of the initial state is $\textbf{u} \in V$. Regarding (i), according to the classical simulation algorithm, the ontic states after one or a larger number of measurements are $\textbf{u}$ or $\textbf{u}+\textbf{a}(Z)$. Either way, the reported measurement outcome is $(-1)^{[\textbf{u},\textbf{a}(Z)]}$, since $[\textbf{a}(Z),\textbf{a}(Z)]=0$. For any sample $\textbf{u}$ from the input distribution, the sequence of measurement outcomes is thus constant, as required. Regarding (ii), since $[\textbf{a}(Z),\textbf{a}(X)]=1$, from the second measurement onwards the outcomes produced by the classical simulation are completely random and uncorrelated, as required.\medskip

{\em{Proof of Theorem~\ref{Simul}.}}  The theorem follows from the efficiency and the correctness of the above classical simulation algorithm.

{\em{Efficiency:}} Both the pre-processing of removing the unitaries from the circuit and the classical simulation algorithm itself need to be considered.

(1) Pre-processing. We need to track  the evolution of Pauli observables $T_\textbf{a}$ under conjugation by gates $g\in G$, as described in Eq.~(\ref{ObsRel}). This can be done efficiently within the stabilizer formalism \cite{Stab}. However, the stabilizer formalism uses its own phase convention $\tilde{\gamma}$ for the Pauli operators, $\tilde{T}_\textbf{a} :=i^{\tilde{\gamma}(\textbf{a})}Z(\textbf{a}_Z)X(\textbf{a}_X)$, such that $\tilde{\gamma}$ can be efficiently evaluated. Suppose, $g^\dagger \tilde{T}_\textbf{a} g = i^{\tilde{\phi}_g(\textbf{a})}\tilde{T}_{g(\textbf{a})}$. Then,  $g^\dagger T_\textbf{a} g = i^{\phi_g(\textbf{a})} T_{g(\textbf{a})}$, with
$$
\phi_g(\textbf{a}) = \tilde{\phi}_g(\textbf{a}) + \left(\tilde{\gamma}(g\textbf{a}) -\tilde{\gamma}(\textbf{a}) \right) -  \left(\gamma(g\textbf{a}) -\gamma(\textbf{a}) \right).
$$
By assumption, $\gamma$ can be efficiently evaluated, and hence can $\phi_g$, for any $g \in G$.

(2) Classical simulation algorithm. The efficiency of the above classical simulation algorithm is evident.\medskip
 
{\em{Correctness:}} Assume that the classical simulation algorithm samples correctly from the Wigner function of the state $\rho_t$ after the $t$-th measurement in the sequence. We now show that under this assumption (i) The above classical simulation algorithm produces the correct probability distribution for the $(t+1)$-th measurement, and (ii) correctly samples from the Wigner function of the conditional state $\rho_{t+1}(s_{t+1})$ after the $(t+1)$-th measurement.

(i) According to the value assignment Eq.~(\ref{ValAss}) of the classical simulation algorithm, the probability $p_\textbf{a}(s)$ for obtaining the outcome $s \in \{0,1\}$ in the measurement of the observable $T_\textbf{a}$ on the state $\rho_t$ is
$$
p_\textbf{a}(s) = \sum_{\textbf{u} \in V}\delta_{[\textbf{a},\textbf{u}],s} W_{\rho_t}(\textbf{u}).
$$ 
As is easily verified by direct calculation,
\begin{equation}\label{EffectW}
W_{\frac{I +(-1)^sT_\textbf{a}}{2}}(\textbf{u}) = \frac{1}{2^n} \delta_{[\textbf{a},\textbf{u}],s}.
\end{equation}
Combining the last two equations, and using the property Eq.~(\ref{Trip}), we find that 
$$
\begin{array}{rcl}
p_\textbf{a}(s) &=& 2^n \sum_{\textbf{u} \in V} W_{\frac{I +(-1)^sT_\textbf{a}}{2}}(\textbf{u}) W_{\rho_t}(\textbf{u})\vspace{1mm}\\
& = &\text{Tr}\left( \frac{I +(-1)^sT_\textbf{a}}{2} \rho_t \right),
\end{array}
$$
which is the quantum-mechanical expression.

(ii) Consider the Wigner function $W_{\rho_t}$ for the state $\rho_t$ after step $t$ in the expansion $W_{\rho_t} = \sum_{\textbf{u}\in V} W_{\rho_t}(\textbf{u}) \delta_\textbf{u}$. At time  $t+1$, the observable $T_\textbf{a}$ is measured, with outcome $s \in \{0,1\}$. Then, from the value assignment Eq.~(\ref{ValAss}), only the phase space points $\textbf{u} \in V$ with $s=[\textbf{u},\textbf{a}]$ contribute to conditional density matrix $\rho_{t+1}(s)$. Furthermore, per Step 2b of the classical simulation algorithm, the update for $\delta$-distributions over phase space is $\delta_\textbf{u} \mapsto (\delta_\textbf{u} +\delta_{\textbf{u} + \textbf{a}})/2$. Therefore, the Wigner function for the (normalized) conditional state $\rho_{t+1}(s)$, according to the classical simulation algorithm, is
$$
p_\textbf{a}(s) W_{\rho_{t+1}(s)} = \sum_{\textbf{u} \in V} \delta_{[\textbf{a},\textbf{u}],s} W_{\rho_t}(\textbf{u}) \frac{\delta_\textbf{u} + \delta_{\textbf{u}+\textbf{a}}}{2}.
$$
Hence, $p_\textbf{a}(s) W_{\rho_{t+1}(s)}(\textbf{v}) = \delta_{[\textbf{a},\textbf{v}],s}\frac{W_{\rho_t}(\textbf{v}) + W_{\rho_t}(\textbf{v}+\textbf{a})}{2}$, which is the quantum-mechanical expression Eq.~(\ref{WigUp}).

By assumption of Theorem~\ref{Simul}, the Wigner function of the initial state $\rho_\text{in} = \rho_0$ is correctly sampled from. Thus, with the above statements (i) and (ii), it follows by induction that all sequences of measurement outcomes occur with the correct probabilities. $\Box$

\subsection{Discussion}\label{Disc}

\noindent
A notable property of the  above simulation method is that, for any Wigner function employed therein, covariance and preservation of positivity under the group $G$ of free gates are not required. This is a consequence of the reformulation of QCSI in Section~\ref{Reform}, where the free unitary gates were eliminated. It is in sharp contrast to the previously considered cases of qudits \cite{NegWi},\cite{Howard} and rebits \cite{ReWi}, where covariance and preservation of positivity under $G$ were critical for the classical simulation by sampling. 
These points, and the roles remaining for covariance and positivity preservation in the present simulation method are discussed below.

\subsubsection{Covariance}

As an example, consider a quantum circuit for a single qubit consisting of a Hadamard gate followed by a measurement of the Pauli observable $Z$. Given is a source that samples from the non-negative Wigner function of the initial state $\rho_\text{in}$, and the task is to sample from the output distribution of the measurement.

A classical simulation method based on Wigner function covariance would, in the first step, convert the source that samples from $W_{\rho_\text{in}}$ into a source that samples from $W_{H\rho_\text{in}H^\dagger}$, using covariance. In the second step, it would, for each sample $\textbf{u}$ drawn from $W_{H\rho_\text{in}H^\dagger}$, output the value $(-1)^{[\textbf{a}(Z),\textbf{u}]}$, with $\textbf{a}(Z)$ such that $T_{\textbf{a}(Z)}=Z$; cf. Eq.~(\ref{ValAss}). But there is a problem: 
\begin{Lemma}\label{NonCov}
For any number $n$ of qubits, no Wigner function of the type defined in Eqs.~(\ref{Tdef}), (\ref{Adef}) is covariant under a Hadamard gate on a single qubit. 
\end{Lemma}
The first step of the above procedure cannot be performed! 

{\em{Proof of Lemma~\ref{NonCov}.}} We only discuss $n=1$, the generalization to other $n$ is straightforward. Consider the phase point operator $A_0= (I+i^{\gamma_x}X+i^{\gamma_y}Y+i^{\gamma_z}Z)/2$. For $W$ to be covariant under $H$, we require that $H^\dagger A_0 H = A_\textbf{u}$, for some $\textbf{u}$. Now consider the sum of signs $\eta = \gamma_x+\gamma_y+\gamma_z \mod 4$, and how it transforms under $H$. Since $H^\dagger X H=Z$, $H^\dagger Y H=-Y$, and $H^\dagger ZH =X$, it follows that $\eta \longrightarrow \eta' =\eta+2 \mod 4$. However, under the transformation $A_0 \longrightarrow A_\textbf{u}=T_\textbf{u}^\dagger A_0 T_\textbf{u}$, the signs of an even number of $\{X,Y,Z\}$ are flipped, hence $\eta$ remains unchanged mod 4. Thus $H^\dagger A_0 H \neq A_\textbf{u}$ for any $\textbf{u}$, for any $\gamma$. $\Box$

\subsubsection{Preservation of positivity}\label{PP}

As an example, consider a quantum circuit for two qubits consisting of a Hadamard gate $H_1$ on the first qubit followed by a measurement of the Pauli observable $Z_1$. Assume the initial state $\rho_\text{in}$ is the completely mixed state, for which each Wigner function of the type Eq.~(\ref{Adef}) is positive and can be efficiently sampled from. The task is to sample from the output distribution of the measurement.

Again, a classical simulation method based on the preservation of Wigner function positivity under free unitaries again runs into a problem: 
\begin{Lemma}\label{NonPosPres}
For $n\geq 2$, for no Wigner function of type Eq.~(\ref{Adef}) positivity is preserved for all states under a Hadamard gate on a single qubit.
\end{Lemma} 

{\em{Proof of Lemma~\ref{NonPosPres}.}} Consider a real stabilizer state $\rho$ of two qubits, $\rho = (I+T_\textbf{a}+T_\textbf{b}+T_\textbf{a}T_\textbf{b})/4$, and w.l.o.g the Hadamard gate $H_1$ on the first qubit. As a consequence of Lemma~\ref{InO}, $W_\rho\geq 0$  if and only if $\beta(\textbf{a},\textbf{b})=0$. Likewise, for the transformed state, $W_{H_1\rho H_1^\dagger}\geq 0$ if and only if $\beta(H_1 \textbf{a}, H_1 \textbf{b})=0$. Now consider the Peres-Mermin square. There are six contexts,  i.e., sets of commuting Pauli observables such that within each set the observables multiply to the identity times $\pm 1$. Whatever the phase convention $\gamma$ for the nine Pauli operators in the square, there is always an odd number of contexts for which $\beta \mod 4=2$. Namely, for the standard phase convention, there is one such context. If the sign of any of the Pauli observables is flipped, then $\beta \longrightarrow \beta + 2 \mod 4$ in the corresponding horizontal and vertical context. Hence the number of contexts with $\beta \mod 4 = 2$ remains odd.

The action of $H_1$ subdivides the set of the six contexts into three orbits of size 2. Since the number of non-zero values of $\beta$ is odd, there must be at least one orbit in which one $\beta$ has the value 0 and the other has the value 2. Within this orbit, choose $\textbf{a},\textbf{b}$ such that $\beta(\textbf{a},\textbf{b})=0$. Hence, $\beta(H_1\textbf{a},H_1\textbf{b})=2$. Thus, positivity is not preserved under $H_1$, for any phase convention $\gamma$. $\Box$

On the other hand, the simulation method of Section~\ref{Reform} has no problem with the above example circuit. Namely, there are Wigner functions of the type Eq.~(\ref{Adef}) for which the Hadamard gate on a single (the first) qubit is in the group of free gates, $H_1 \in G$, and $Z_1$ is in the set of directly measurable observables, $Z_1 \in {\cal{O}}$. An example for such a Wigner function is given in Section~\ref{qubitQCSI}.

We observe that the negativity which can be introduced into a Wigner function by the free unitary gates $G$ is of a very special kind. Namely, it can be lifted by redefinition of the Wigner function according to $A_\textbf{v} \mapsto A_\textbf{v}' = g A_\textbf{v} g^\dagger$, $\forall \textbf{v} \in V$, for some $g \in G$.

To summarize, while in the present framework the free measurements are required to preserve positivity of the Wigner function, no such constraint needs to be imposed on the free unitaries. {The amount of negativity introduced into the Wigner functions by the free unitaries can be large, as measured by sum negativity \cite{EmRes}. However, it is always of a special kind.  In this sense, our observation complements the recent finding \cite{Bartl} that a small amount of sum negativity---of any kind---does not compromise the efficiency of a suitable classical simulation algorithm.}

\section{Contextuality}\label{Context}

In this section we establish contextuality of magic states as a computational resource, for qubit QCSI schemes satisfying the condition (P1). We also clarify how contextuality and Wigner function negativity are related, and state our most general classical simulation algorithm for qubit QCSI.

\subsection{Non-contextual hidden variable models}\label{HVMdef}

Recall that ${\cal{O}}$ is the set of Pauli observables which can be directly measured in QCSI, $M$ is the set of observables which can have their value inferred by measurement of observables in ${\cal{O}}$, and any $C \subset M$ is a set of Pauli observables which can have their value inferred jointly, from a single copy of the given quantum state.

\begin{Def}\label{HVM1} Consider a quantum state $\rho$ and a set ${\cal{O}}$ of observables grouping into contexts $C$ of simultaneously measurable observables.
 A non-contextual hidden variable model $({\cal{S}}, q_\rho,\Lambda)$  consists of a probability distribution $q_\rho$ over a set ${\cal{S}}$ of internal states and a set $\Lambda=\{\lambda_\nu\}_{\nu\in{\cal{S}}}$ of value assignment functions  $\lambda_\nu: {\cal{O}}\rightarrow\mathbb{R}$ that meet the following criteria.
 \begin{itemize}
  \item[(i)]  Each $\lambda_\nu\in\Lambda$ is consistent with quantum mechanics: for any set $C$ of jointly measurable observables there exists a quantum state $|\psi\rangle$ such that
  \begin{equation}\label{eq:JointEigenvalue}
  A|\psi\rangle=\lambda_\nu(A)|\psi\rangle, \quad\forall A\in C.
  \end{equation}
   \item[(ii)] The distribution $q_\rho$ satisfies
   \begin{equation}\label{Aexpct}
   \mathrm{tr}(A \rho)=\sum_{\nu\in\mathcal{S}} \lambda_\nu(A)q_\rho(\nu), \quad \forall A\in M.
   \end{equation}
  \end{itemize}
 \end{Def}
We say that a quantum state $\rho$ is contextual if no non-contextual HVM according to Def.~\ref{HVM1} exists that correctly reproduces the probability distributions $p_{C,\rho}(\textbf{s}_C)$ of measurement outcomes for all sets $C$ of jointly measurable observables.

{The states $|\psi\rangle$ in Eq.~(\ref{eq:JointEigenvalue}) of Def.~\ref{HVM1} are auxiliary. Their purpose is to ensure that the value assignments $\lambda_\nu$ correspond to compatible eigenvalues. As a direct consequence of Eq.~(\ref{eq:JointEigenvalue}), the non-contextual value assignments $\lambda_\nu \in \Lambda$ must all satisfy a set of compatibility constraints.
\begin{Lemma}\label{ProdConstr}
 For any triple $A,B,AB \in M$ of simultaneously measurable observables and any internal state $\nu \in {\cal{S}}$ of an NCHVM   $({\cal{S}}, q_\rho,\Lambda)$ it holds that
 \begin{equation}\label{Valco}
 \lambda_\nu(AB) =  \lambda_\nu(A)  \lambda_\nu(B).
 \end{equation}
 \end{Lemma}}
{
Returning to Def.~\ref{HVM1}, it may {\em{a priori}} seem that the condition (ii) is not sufficiently stringent, and that one should rather require all outcome distributions for sets $C$ of jointly measurable observables to be correctly reproduced by the hidden variable model. That is,
\begin{equation}\label{Cprob}
p_{C,\rho}(\textbf{s}_C) =\sum_{\nu \in {\cal{S}}} p(\textbf{s}_C|\nu)q_\rho(\nu).
\end{equation}
Therein, $p_{C,\rho}$ is the probability distribution for the measurement outcomes $\textbf{s}_C$ of the set $C$ of simultaneously measurable observables given the quantum state $\rho$, and $p(\textbf{s}_C|\nu)$ is the conditional probability for the measurement outcomes $\textbf{s}_C$ given the HVM internal state $\nu$. However, Eq.~(\ref{Cprob}) is implied by Eq.~(\ref{Aexpct}).}

\begin{Lemma}\label{probs}
An ncHVM according to Def.~\ref{HVM1} that correctly reproduces all expectation values of observables $A\in M$ via Eq.~(\ref{Aexpct}) also correctly reproduces the outcome probability distributions Eq.~(\ref{Cprob}) for all sets $C\subset M$ of jointly measurable observables. 
\end{Lemma}

{\em{Proof of Lemma~\ref{probs}.}} Assume the observables in $C$ are algebraically independent, i.e., there are no non-trivial product relations among them, and denote by $\text{span}(C)$ the set of all products of observables in $C$.

{Quantum mechanically,
$p_{C,\rho}(\textbf{s}_C) = \text{Tr}(E_C(\textbf{s}_C)\rho)$, where the effect $E_C(\textbf{s}_C)$ is
$$
\begin{array}{rcl}
E_C(\textbf{s}_C)&:=& \displaystyle{\prod_{A\in C} \frac{I+(-1)^{s(A)}A}{2}}\vspace{1mm}\\
&=& \displaystyle{\frac{1}{2^{|C|}} \sum_{B \in \text{span}(C)}(-1)^{s(B)}B.}
\end{array}
$$
Therein, in the last line we have used that the measured eigenvalues $(1)^{s(B)}$ satisfy the same consistency relation Eq.~(\ref{Valco}) as the non-contextual value assignments. Then,
$$
\begin{array}{rcl}
p_{C,\rho}(\textbf{s}_C) &=& \displaystyle{\frac{1}{2^{|C|}}\sum_{B \in \text{span}(C)}(-1)^{s(B)} \text{Tr} (B\rho)}\\
&=& \displaystyle{\frac{1}{2^{|C|}}\sum_{B \in \text{span}(C)}(-1)^{s(B)}\sum_{\nu\in {\cal{S}}} \lambda_\nu(B) q_\rho(\nu)}\\
&=& \displaystyle{\frac{1}{2^{|C|}} \sum_{\nu\in {\cal{S}}} q_\rho(\nu) \sum_{B \in \text{span}(C)}(-1)^{s(B)} \lambda_\nu(B) }\\
&=& \displaystyle{\sum_{\nu\in {\cal{S}}} q_\rho(\nu) \delta_{(-1)^{\textbf{s}_C},\lambda_\nu|_C}}\\
&=& \displaystyle{\sum_{\nu\in {\cal{S}}} q_\rho(\nu) p(\textbf{s}_C|\nu)}.
\end{array}
$$
In the last line above, we have used the fact that the value assignments $\lambda$ are deterministic and that the conditional probabilities $p(\textbf{s}_C|\nu)$ are thus $\delta$-functions. $\Box$
}

\subsection{The absence of state-independent contextuality}\label{NoSIC}

Consider the value assignment 
\begin{equation}\label{1}
\lambda(T_\textbf{a}) =1, \;\; \forall \textbf{a} \in V_M.
\end{equation}
First, by Eq.~(\ref{C1}), for any three commuting and directly measurable observables $T_\textbf{a}, T_\textbf{b},T_{\textbf{a} + \textbf{b}} \in {\cal{O}}$ we have $T_{\textbf{a}+\textbf{b}} = + T_\textbf{a} T_\textbf{b}$. Thus, the above value assignment is compatible with all available direct measurements. 

Second, the value of any observable $T_{\textbf{a}+\textbf{b}}\in M\backslash {\cal{O}}$ is inferred by measuring a suitable observable $T_\textbf{a} \in {\cal{O}}$ for which $[T_{\textbf{a}+\textbf{b}},T_\textbf{a}]=0$, and then running a procedure to infer the value of $T_\textbf{b}$. With Eq.~(\ref{C1}), for all such observables $T_\textbf{a}$ it holds that $T_{\textbf{a}+\textbf{b}} = + T_\textbf{a} T_\textbf{b}$, and the assignment Eq.~(\ref{1}) is thus consistent.

While the Peres-Mermin square and its cousins are present, the operational restriction enforced by condition Eq.~(\ref{C1}) prevents obstructions to the assignment Eq.~(\ref{1}) from being established as experimental facts. Hence at least one consistent assignment exists, and there is no state-independent contextuality in this setting.\medskip

We are now in the position to prove Lemma~\ref{C+} of Section \ref{ConsCond}.

{\em{Proof of Lemma~\ref{C+}.}} If there is a set $C$ with $T_\textbf{a},T_\textbf{b} \in C$ such then $T_{\textbf{a}+\textbf{b}} = -T_\textbf{a} T_\textbf{b}$ then the value assignment $\lambda(T_\textbf{a}) =1$,  $\forall \textbf{a} \in V$, is inconsistent. Contradiction. $\Box$\medskip

{{\em{Remark:}} In the companion paper \cite{BBW}, the condition (P1) is replaced by the requirement that the set $M$ of observables with inferable values is free of state-independent contextuality, given the set ${\cal{O}}$ of directly measurable observables. With the above, we find that all such QCSI schemes are included in the present classification. The converse also holds: for every qubit QCSI scheme in which the set $M$ of observables is free of state-independent contextuality, there is a Wigner function $W^\gamma$ such that the condition (P1) holds.}

{This can be seen as follows. If $M$ is free of state-independent contextuality given the set ${\cal{O}}$ of directly measurable observables, then there exists at least one consistent value assignment $\lambda: V_M \longrightarrow \{\pm 1\}$. We may now re-phase the inferable observables, $T_\textbf{a} \mapsto T'_\textbf{a} = \lambda(\textbf{a})^{-1}T_\textbf{a}$, for all $\textbf{a} \in V_M$, such that the new observables $\{T'_\textbf{a}, \textbf{a}\in V_M\}$ have a consistent value assignment $\lambda'\equiv 1$. This means that for all $T'_\textbf{a} \in {\cal{O}}$, $T'_\textbf{b},T'_{\textbf{a}+\textbf{b}} \in M$ with $[\textbf{a},\textbf{b}]=0$ it holds that $T'_{\textbf{a}+\textbf{b}} = + T'_\textbf{a}T'_\textbf{b}$. With Eq.~(\ref{3T}) it thus follows that $\beta'(\textbf{a},\textbf{b})=0$ for all $\textbf{a} \in V_{\cal{O}}$, $\textbf{b} \in V_M$ with $[\textbf{a},\textbf{b}]=0$. Thus, with Lemma~\ref{InO}, the measurement of any observable $\pm T'_\textbf{a}$ does not introduce negativity into the Wigner function defined by $A_0=1/2^n\sum_{\textbf{a}\in V}T'_\textbf{a}$. 
}

\subsection{Contextuality implies Wigner negativity}\label{NWNC}

In accordance with existing results \cite{Spekk}, \cite{Howard},  also for the present setting of QCSI on qubits it holds that a non-negative Wigner function always implies the viability of a non-contextual hidden variable model.

\begin{Theorem}\label{pwnc}
Consider a quantum state $\rho$ with Wigner function $W^\gamma_\rho$ given by Eq.~(\ref{Adef}). If $W^\gamma_\rho \geq 0$ then the measurement of all Pauli observables $T_\textbf{a} \in {\cal{O}}$ can be described by a non-contextual hidden variable model.
\end{Theorem}

{\em{Proof of Theorem~\ref{pwnc}.}} The Wigner function itself constitutes a non-contextual HVM, with set of internal states ${\cal{S}}=V$, probability distribution $q_\rho(\textbf{u})=W_\rho(\textbf{u})$ over the internal states, and the conditional probabilities $p(\textbf{s}_C|\textbf{u})$ given by the Wigner functions of the effects. 

Using Eq.~(\ref{Trip}), the probability $p_C(\textbf{s}_C)$ for measuring the context $C$ and obtaining the set $\textbf{s}_C$ of measurement outcomes is
\begin{equation}\label{Bayes}
p_C(\textbf{s}_C) = \text{Tr}(\rho E_C(\textbf{s}_C)) = \sum_{\textbf{u}\in V} W_\rho(\textbf{u}) (2^n W_{E_C(\textbf{s}_C)}(\textbf{u})).
\end{equation}
If $W_\rho \geq 0$ then $W_\rho$ may be regarded as a probability distribution over the space $V$ of internal states of a hidden variable model. If furthermore $0 \leq 2^n W_{E_C(\textbf{s}_C)}(\textbf{u}) \leq 1$ for all $\textbf{u} \in V$, then we may regard $2^n W_{E_C(\textbf{s}_C)}=:p(\textbf{s}_C|\textbf{u})$ as the conditional probability for obtaining the outcome $\textbf{s}_C$ in the measurement of the observables in $C$, given the HVM internal state $\textbf{u} \in V$. Then, Eq.~(\ref{Bayes}) is Bayes' rule for computing the probability $p_C(\textbf{s}_C)$. 

{It remains to check that the conditional probabilities $p(\textbf{s}_C|\textbf{u})$ assigned by the Wigner functions of the effects $E_C(\textbf{s}_C)$ are compatible with Def.~\ref{HVM1} of an ncHVM.}

{Item (i) of Def.~\ref{HVM1}. Consider contexts wit a single observable $T_\textbf{a}$.} A definite value assignment for all internal states $\textbf{u} \in V$ has already been established in Eq.~(\ref{ValAss}). It corresponds to 
\begin{equation}\label{EffWi}
p_\textbf{a}(s|\nu) =  2^n W_{E_\textbf{a}(s)}(\textbf{u}) = \delta_{s,[\textbf{a},\textbf{u}]}.
\end{equation}
The conditional probabilities $p_\textbf{a}(s|\nu)$ thus have values in the required range $0 .. 1$. 

{Consistency of the value assignments. Eq.~(\ref{ValAss}) yields $\lambda_\textbf{u}(T_{\textbf{a}+\textbf{b}})= \lambda_\textbf{u}(T_\textbf{a})\lambda_\textbf{u}(T_\textbf{b})$, for all $\textbf{a},\textbf{b} \in V$. By Lemma~\ref{C+}, if $\{T_\textbf{a},T_\textbf{b}\}$ form a jointly measurable set, then $T_{\textbf{a}+\textbf{b}}=T_\textbf{a}T_\textbf{b}$. Hence,  $\lambda_\textbf{u}(T_\textbf{a}T_\textbf{b}) = \lambda_\textbf{u}(T_\textbf{a}) \lambda_\textbf{u}(T_\textbf{b})$ for such pairs $\textbf{a},\textbf{b} \in V$, as required by Lemma~\ref{ProdConstr}. Therefore, for all contexts $C$, the observables $T_\textbf{a} \in C$ have a joint eigenstate $|\psi_0\rangle$ with $T_\textbf{a}|\psi_0(C)\rangle = +|\psi_0(C)\rangle$, and thus, the state $|\psi_\textbf{u}(C)\rangle := T_\textbf{u} |\psi_0(C)\rangle$ has eigenvalues $(-1)^{[\textbf{a},\textbf{u}]}$ for the observbles $T_\textbf{a} \in C$. The condition Eq.~(\ref{eq:JointEigenvalue}) is thus satisfied for the value assignment Eq.~(\ref{ValAss}).}  

{Item (ii) of Def.~\ref{HVM1}. Since all observables $T_\textbf{a} \in M$ only have eigenvalues $\pm 1$,
$$
\begin{array}{rcl}
\text{Tr}(T_\textbf{a}\rho) &=& \displaystyle{\text{Tr}\left((E_\textbf{a}(0)-E_\textbf{a}(1))\rho\right)}\vspace{1mm}\\
&=& \displaystyle{2^n \sum_{\textbf{u}\in V} W_\rho(\textbf{u}) \left(W_{E_\textbf{a}(0)}-W_{E_\textbf{a}(1)}\right)}\vspace{1mm}\\
&=& \displaystyle{\sum_{\textbf{u}\in V} W_\rho(\textbf{u}) \left(\delta_{0,[\textbf{a},\textbf{u}]}-\delta_{1,[\textbf{a},\textbf{u}]}\right)}\vspace{1mm}\\
&=& \displaystyle{\sum_{\textbf{u}\in V} W_\rho(\textbf{u}) (-1)^{[\textbf{a},\textbf{u}]}}\vspace{1mm}\\
&=& \displaystyle{\sum_{\textbf{u}\in V} q_\rho(\textbf{u})\lambda_\textbf{u}(\textbf{a})}.
\end{array}
$$
Above, the second line follows by Eq.~(\ref{Trip}), the third by Eq.~(\ref{EffWi})  and the fifth by Eq.~(\ref{ValAss}). $\Box$}
\medskip

\begin{figure}
\begin{center}
\includegraphics[width=7.5cm]{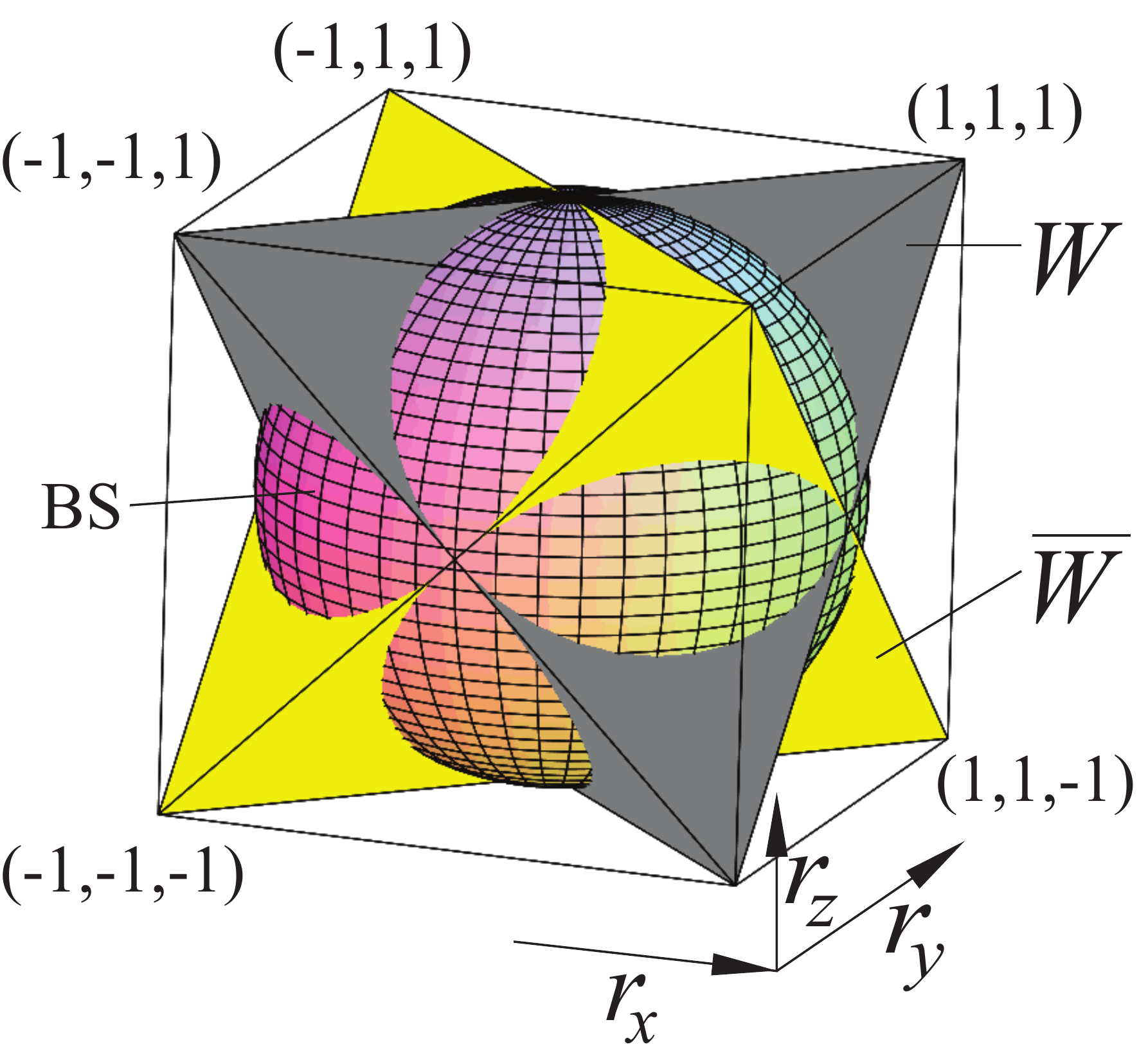}
\caption{\label{QubitHVM} State space for the one-qubit states $\rho = (I+\textbf{r}\vec{\sigma})/2$. The physical states lie within or on the Bloch sphere (BS). The two tetrahedra contain the states positively represented by the Wigner functions $W$ and $\overline{W}$, respectively. The state space describable by a non-contextual HVM is a cube with corners $(\pm 1,\pm 1,\pm 1)$; also see \cite{Bartl2}. It contains the Bloch ball.}
\end{center}
\end{figure}

The converse of Theorem~\ref{pwnc} does not hold: there are quantum states with a non-contextual HVM description for which all considered Wigner functions are negative. This is  illustrated in Fig.~\ref{QubitHVM} for the example of a single qubit, where all physically allowed states have an HVM description \cite{Bell}. The one-qubit states are all of the form $\rho=(I+\textbf{r}\vec{\sigma})/2$, and the physical such states are constrained by $|\textbf{r}|\leq 1$. The set of states describable in terms of a non-contextual HVM is a cube, $|r_x|,|r_y|,|r_z| \leq 1$, containing all physical states. The eight extremal states $i$ of this cube have definite value assignments $\lambda_i(X),\lambda_i(Y),\lambda_i(Z) = \pm 1$ for the observables $X$, $Y$, $Z$.

Up to equivalence under translation, there are two one-qubit Wigner functions of type Eq.~(\ref{Adef}), namely the Wigner function $W$ defined by the phase point operator at the origin $A_\textbf{0}= (I+X+Z+Y)/2$, and the Wigner function $\overline{W}$ defined by $\overline{A}_\textbf{0}=HA_\textbf{0}H^\dagger = (I+X+Z-Y)/2$. The phase space for these Wigner functions is $\mathbb{Z}_2\times \mathbb{Z}_2$, and $W$, $\overline{W}$ thus have four extremal states each. If these extremal states are combined, the extremal states of the non-contextual HVM are recovered. Each Wigner function by itself has only half of the extremal states of the HVM, and the set of positively represented states is thus smaller. Furthermore, there are physical states which are negatively represented by both $W$ and $\overline{W}$; See Fig.~\ref{QubitHVM}. Contextuality and negativity of the Wigner functions Eq.~(\ref{Adef}) are thus not the same. \medskip

{\em{Remark:}} While there are physical one-qubit quantum states which are negatively represented by both $W$ and $\overline{W}$, every state $\rho=(I+\textbf{r}\vec{\sigma})/2$, with $|r_x|,|r_y|,|r_z|\leq 1$, can be described by an ensemble 
$$
{\cal{E}}(\rho) =\{(p_1,\rho_1), (p_2,\rho_2)\}, 
$$
such that $W_{\rho_1}\geq 0$ and $\overline{W}_{\rho_2} \geq 0$. A generalization of this fact to $n$-qubit systems will be of relevance in Section~\ref{CR}.

\subsection{Contextuality as a resource}\label{CR}

\begin{Theorem}\label{Res1}
{For any QCSI scheme $(\gamma, {\cal{M}})$}, if the input magic state $\rho_\text{in}$ can be described by a non-contextual HVM, then the quantum state $\rho_t(\textbf{s}_{\prec t})$ at time $t$, conditioned on the prior measurement record $\textbf{s}_{\prec t}$, can be described by a non-contextual HVM, for any $t$ and any $\textbf{s}$. 
\end{Theorem}
Before we give the proof of Theorem~\ref{Res1}, we need to set up some more notation. We observe that the state space of a general non-contextual HVM is larger than the state space of an HVM deriving from a non-negative Wigner function; See the discussion of a single qubit in Section~\ref{NWNC}/ Fig.~\ref{QubitHVM}. The enlarged state space ${\cal{S}}=\{\nu\}$ is finite, yet maximal in the sense that, for every value assignment $\lambda(\cdot)$ satisfying the consistency condions Eq.~(\ref{Valco}), there is a corresponding internal state $\nu \in {\cal{S}}$ such that $\lambda_\nu(\cdot) \equiv \lambda(\cdot)$.

We choose to have this state space ${\cal{S}}$ acted upon by the group $V$, dividing ${\cal{S}}$ into orbits. Namely, given an element $\nu \in {\cal{S}}$ specified by the value assignment $\lambda_\nu: V \longrightarrow \{\pm1\}$, there is another internal state $\nu +\textbf{u}$, defined through the value assignment
\begin{equation}\label{Transl}
\lambda_{\nu +\textbf{u}}(\textbf{a}) = \lambda_\nu(\textbf{a}) (-1)^{[\textbf{u},\textbf{a}]},\;\forall \textbf{a} \in V,
\end{equation}
for all $\textbf{u} \in V$. In Eq.~(\ref{Transl}), we have set $\lambda_\nu(\textbf{a}):=\lambda_\nu(T_\textbf{a})$ for notational simplicity. It is easily seen that $\nu \in {\cal{S}} \Leftrightarrow \nu+\textbf{u} \in {\cal{S}}$, for all $\textbf{u} \in V$.  The condition to check is the consistency of the value assignment in item (ii) of Def.~\ref{HVM1}. Eq.~(\ref{Valco}) is preserved under the change $\lambda_\nu(\textbf{a}) \mapsto \lambda_\nu(\textbf{a}) (-1)^{[\textbf{u},\textbf{a}]}$, for any $\textbf{u} \in V$. The group action of $V$ on ${\cal{S}}$ defined through Eq.~(\ref{Transl}) labels the elements of ${\cal{S}}$ in a fashion convenient for the subsequent discussion.
\medskip

{\em{Proof of Theorem~\ref{Res1}.}} 
The proof of Theorem~\ref{Res1} is by induction. We assume that there exists an HVM with probability distribution $q_{t,\textbf{s}_{\prec t}}$ which describes the quantum state $\rho_{t}(\textbf{s}_{\prec t})$, conditioned on the previous measurement record $\textbf{s}_{\prec t}$. We then show that there is an HVM with probability distribution $q_{t+1,\textbf{s}_{\prec t+1}}$ which describes the quantum state $\rho_{t+1}(\textbf{s}_{\prec t+1})$. 

To establish this result, we need the relation between $q_{t+1,\textbf{s}_{\prec t+1}}$ and its precursor $q_{t,\textbf{s}_{\prec t}}$. Denoting the observable measured in the $t$-th time step of the computation by $T_{\textbf{a}_t}\in {\cal{O}}$ and the corresponding measurement outcome by $s_t \in \mathbb{Z}_2$, the required  relation is 
\begin{subequations}\label{HVMrelat}
\begin{align}
\label{HVMq}
q_{t+1,\textbf{s}_{\prec t+1}}(\nu) &= \displaystyle{\frac{\delta_{(-1)^{s_t},\lambda_\nu(\textbf{a}_t)}}{\overline{p}_t(s_t|\textbf{s}_{\prec t})}   \frac{q_{t,\textbf{s}_{\prec t}}(\nu) +  q_{t,\textbf{s}_{\prec t}}(\nu +\textbf{a}_t)}{2},}\vspace{2mm}\\
\label{HVMp}
\overline{p}_t(s_t|\textbf{s}_{\prec t}) &= \displaystyle{\sum_{\nu \in {\cal{S}}} \delta_{(-1)^{s_t},\lambda_\nu(\textbf{a}_t)} q_{t,\textbf{s}_{\prec t}}(\nu).}
\end{align}
\end{subequations}
Therein, $\overline{p}_t(s_t|\textbf{s}_{\prec t})$ is the HVM prediction for the probability of obtaining the outcome $s_t$ in the measurement of $T_{\textbf{a}_t}$, given a prior measurement record $\textbf{s}_{\prec t}$. Eq.~(\ref{HVMrelat}) will be justified a posteriori. Namely, with these assignments, the induction argument works out.

With Eq.~(\ref{Aexpct}) in Def.~\ref{HVM1}, the induction assumption is
$$
\langle T_\textbf{a} \rangle_{\rho_{t}} = \langle T_\textbf{a}\rangle_{q_t},\;\forall \textbf{a} \in V.
$$
Therein, we have suppressed the dependence on the measurement record, to simplify the notation. We need to show that
$$
\langle T_\textbf{a} \rangle_{\rho_{t+1}} = \langle T_\textbf{a}\rangle_{q_{t+1}},\;\forall \textbf{a} \in V,
$$
and that $\overline{p}_t(s_t|\textbf{s}_{\prec t}) = p_t(s_t|\textbf{s}_{\prec t})$, with $p_t(s_t|\textbf{s}_{\prec t})$ the quantum mechanical value for the probability of the outcome $s_t$ given the prior measurement record $\textbf{s}_{\prec t}$.

First, regarding the probability of finding $s_t$,
$$
\begin{array}{rcl}
\overline{p}_t(s_t|\textbf{s}_{\prec t}) &=& \displaystyle{\sum_{\nu \in {\cal{S}}} \frac{1+(-1)^{s_t}\lambda_\nu(\textbf{a}_t)}{2} q_{t,\textbf{s}_{\prec t}}(\nu)}\vspace{1mm}\\
&=& \displaystyle{\frac{1}{2}\sum_{\nu\in {\cal{S}}} q_{t,\textbf{s}_{\prec t}}(\nu) +  \frac{(-1)^{s_t}}{2}\sum_{\nu\in {\cal{S}}} q_{t,\textbf{s}_{\prec t}}(\nu) \lambda_\nu(\textbf{a}_t)}\vspace{1mm}\\
&=& \displaystyle{\frac{\langle I\rangle_{\rho_t(\textbf{s}_{\prec t})} +  (-1)^{s_t}   \langle T_{\textbf{a}_t} \rangle_{\rho_t(\textbf{s}_{\prec t})}   }{2}}\vspace{1mm}\\
&=& \displaystyle{\text{Tr} \left( \rho_t(\textbf{s}_{\prec t}) \frac{I+(-1)^{s_t} T_{\textbf{a}_t}}{2} \right)}\vspace{1mm}\\
&=& \displaystyle{p_t(s_t|\textbf{s}_{\prec t}).}
\end{array}
$$
We thus reproduce the quantum mechanical expression within the HVM.
Above, in transitioning from the second to the third line we have invoked the induction assumption.

Second, regarding the expectation values of the $T_\textbf{a}$ on $\rho_{t+1}(\textbf{s}_{\prec t+1})$, the HVM prediction is
$$
\begin{array}{rcl}
\langle T_\textbf{a} \rangle_{q_{t+1}} &=& \displaystyle{{\sum_{\nu \in {\cal{S}}}\!} q_{t+1,\textbf{s}_{\prec t+1}}(\nu) \lambda_\nu(\textbf{a})} \vspace{1mm}\\
&=& \displaystyle{\sum_{\nu \in {\cal{S}}}\! \frac{1+(-1)^{s_t}\lambda_\nu(\textbf{a}_t)}{4 p_t(s_t|\textbf{s}_{\prec t})} q_{t,\textbf{s}_{\prec t}}(\nu)   \lambda_\nu(\textbf{a})+}\\
&& \displaystyle{+\sum_{\nu \in {\cal{S}}}\! \frac{1\!+\!(-1)^{s_t}\lambda_\nu(\textbf{a}_t)}{4 p_t(s_t|\textbf{s}_{\prec t})} q_{t,\textbf{s}_{\prec t}}(\nu \! + \! \textbf{a}_t)   \lambda_\nu(\textbf{a}).}
\end{array}
$$
Reordering the sum via the substitution $\nu + \textbf{a}_t \rightarrow \nu$, and using Eq.~(\ref{Transl}), the second term in the last line equals
$$
\displaystyle{\sum_{\nu \in {\cal{S}}}\! \frac{1\!+\!(-1)^{s_t}\lambda_\nu(\textbf{a}_t)}{4 p_t(s_t|\textbf{s}_{\prec t})} q_{t,\textbf{s}_{\prec t}}(\nu \! )   \lambda_\nu(\textbf{a}) (-1)^{[\textbf{a}, \textbf{a}_t]}.}
$$
We now distinguish between the case where $T_\textbf{a}, T_{\textbf{a}_t}$ commute and where they don't.

Case (i): $[\textbf{a},\textbf{a}_t]=1$. Then, $\langle T_\textbf{a} \rangle_{q_{t+1}}=0$, which is the correct quantum mechanical expression.

Case (ii): $[\textbf{a},\textbf{a}_t]=0$. Then, the expression for $\langle T_\textbf{a} \rangle_{q_{t+1}}$ simplifies to
$$
\begin{array}{rcl}
\langle T_\textbf{a} \rangle_{q_{t+1}} &=& \displaystyle{\sum_{\nu \in {\cal{S}}}\! \frac{1+(-1)^{s_t}\lambda_\nu(\textbf{a}_t)}{2 p_t(s_t|\textbf{s}_{\prec t})} q_{t,\textbf{s}_{\prec t}}(\nu)   \lambda_\nu(\textbf{a})}\\
&=& \displaystyle{\frac{1}{2p_t(s_t|\textbf{s}_{\prec t})} \sum_{\nu \in {\cal{S}}} q_{t,\textbf{s}_{\prec t}}(\nu)  \lambda_\nu(\textbf{a})+}\\
&& \displaystyle{+\frac{(-1)^{s_t}}{2p_t(s_t|\textbf{s}_{\prec t})} \sum_{\nu \in {\cal{S}}} q_{t,\textbf{s}_{\prec t}}(\nu)  \lambda_\nu(\textbf{a}+\textbf{a}_t).}
\end{array}
$$
Here we have used the relation $\lambda_\nu(\textbf{a}+\textbf{a}_t)=\lambda_\nu(\textbf{a}_t)\lambda_\nu(\textbf{a})$, which arises as follows. Since $T_{\textbf{a}_t} \in {\cal{O}}$, $T_\textbf{a} \in M$, and $[T_{\textbf{a}_t},T_\textbf{a}]=0$ by the case assumption, $\{T_{\textbf{a}_t},T_\textbf{a}\}$ is a jointly measurable set of observables; cf. example (iii) after Def.~\ref{DefC}. (The procedure is to measure $T_{\textbf{a}_t} \in {\cal{O}}$ first, and then run the measurement sequence for $T_\textbf{a} \in M$.) Thus, by Property (ii) of Def.~\ref{HVM1} for non-contextual HVMs, $\lambda_\nu(T_{\textbf{a}_t}T_\textbf{a})=\lambda_\nu(T_{\textbf{a}_t})\lambda_\nu(T_{\textbf{a}})$. Finally, with Lemma~\ref{InO}, $T_{\textbf{a}_t}T_\textbf{a}=T_{\textbf{a}+\textbf{a}_t}$, which yields the stated relation.

Next we use the induction assumption, and obtain
$$
\begin{array}{rcl}
\langle T_\textbf{a} \rangle_{q_{t+1}} &=&\displaystyle{\frac{1}{2p_t(s_t|\textbf{s}_{\prec t})} \left( \langle T_\textbf{a}\rangle_{\rho_t} (-1)^{s_t}  + \langle T_{\textbf{a}+\textbf{a}_t}\rangle_{\rho_t} \right)}\vspace{1mm}\\
&=& \displaystyle{\frac{1}{p_t(s_t|\textbf{s}_{\prec t})} \text{Tr} \left( \rho_t \frac{I+(-1)^{s_t} T_{\textbf{a}_t}}{2} T_\textbf{a} \right)}\vspace{1mm}\\
&=& \displaystyle{\frac{\text{Tr} \left( \left[\frac{I+(-1)^{s_t} T_{\textbf{a}_t}}{2} \rho_t \frac{I+(-1)^{s_t} T_{\textbf{a}_t}}{2} \right] T_\textbf{a} \right)}{p_t(s_t|\textbf{s}_{\prec t})}}\vspace{1mm}\\
&=& \displaystyle{\langle T_\textbf{a} \rangle_{\rho_{t+1}}.}
\end{array}
$$
We thus reproduce the quantum mechanical expression within the HVM. This completes the induction step. 

The induction starts at time $t=1$, where $\rho_1=\rho_\text{in}$ has an HVM description, by assumption of Theorem~\ref{Res1}. Thus, by induction, for every time $t\geq 1$ and every history $\textbf{s}_{\prec t}$ of measurement outcomes, the conditional state $\rho_t(\textbf{s}_{\prec t})$ has a description in terms of a non-contextual HVM. $\Box$

\begin{Cor}\label{ProbPredict}
For any QCSI scheme $(\gamma, {\cal{M}})$, if the input magic state $\rho_\text{in}$ can be described by a non-contextual HVM, then for the measurement of any sequence of observables $\{T_{\textbf{a}_t},t =1.. t_\text{max}\}  \subset {\cal{O}}$ , the probability distribution $p(\textbf{s})=p(s_1,s_2,..,s_{t_\text{max}})$ of outcomes is fixed by the HVM for $\rho_\text{in}$. The $T_{\textbf{a}_t}$ may be mutually non-commuting and dependent on previous measurement outcomes.
\end{Cor}

{\em{Proof of Corollary~\ref{ProbPredict}.}} By Bayes' rule, the joint probability of the outcomes $\textbf{s}$ can be written as
$$
\begin{array}{rcl}
p(\textbf{s}) &=&  \displaystyle{\prod_{t=1}^{t_\text{max}} p_t(s_t|\textbf{s}_{\prec t})}.
\end{array}
$$
By Theorem~\ref{Res1}, the conditional probabilities $p_t(s_t|\textbf{s}_{\prec t})=\overline{p}_t(s_t|\textbf{s}_{\prec t})$ are all correctly obtained from the probability distributions $q_{t,\textbf{s}_{\prec t}}$, cf. Eq.~(\ref{HVMp}).  The distributions $q_{t,\textbf{s}_{\prec t}}$, for $t=2,..,t_\text{max}$, in turn follow from the distribution $q_{1,\textbf{s}_{\prec 1}=\emptyset}$ (describing $\rho_\text{in}$ at $t=1$), by Eq.~(\ref{HVMq}). Thus, $p(\textbf{s})$ is fully specified by the probability distribution  $q_{1,\textbf{s}_{\prec 1}=\emptyset}$ over the state space ${\cal{S}}$ of the HVM. $\Box$\medskip

{We now discuss the implications of Theorem~\ref{Res1} with regards to universality of quantum computation. We want to capture in our analysis the case where a QCSI scheme running on $n$ qubits is universal only on a subspace supporting $k$ encoded qubits. (This does of course include the unencoded case, where every logical qubit is represented by one physical qubit.) We use the following notion of computational universality.
\begin{Def}
We say that a QCSI scheme is encoded universal if the following operations can be performed.
\begin{itemize}
\item[U1]{{\em{Encoded inputs.}} Prepare a  set of encoded orthonormal input states ${\cal{E}}(|x\rangle),\; x\in\{0,1\}^k$ up to an arbitrarily small error $\epsilon$,  where ${\cal{E}}: \mathbb{C}^{2^k}\longrightarrow \mathbb{C}^{2^n}$ is an isometry of $k$ logical qubits into $n$ physical qubits.}
\item[U2]{{\em{Encoded gates.}} For any $V \in SU (\mathbb{C}^{2^k})$ and any encoded input state ${\cal{E}}(|\phi\rangle)$ prepare the encoded output state ${\cal{E}}(V|\phi\rangle)$, up to an arbitrarily small error $\epsilon$.}
\item[U3]{{\em{Encoded outputs}}. Measure the value of any logical observable ${\cal{E}}(X_i)$, i.e., $\{{\cal{E}}(X_i),\; i=1,..,k\} \subset {\cal{O}}$.}
\end{itemize}
\end{Def}
Requirement U3 means that it is possible to physically measure any logical qubit in the standard basis.}

{We then have the following result.
\begin{Theorem}\label{UnivCon}
A QCSI scheme $(\gamma,{\cal{M}})$ on $k\geq 3$ (possibly encoded) qubits satisfying U1 - U3 is universal only if its magic states are contextual.
\end{Theorem}
The full proof of Theorem~\ref{UnivCon} is given in Appendix~\ref{UnivCP}. Here we prove Theorem~\ref{UnivCon} under the simplifying assumption that every encoded qubit can be measured in two complementary bases rather than one basis. That is, U3 is replaced by 
\begin{itemize}
\item[U$3'$]{$\{{\cal{E}}(X_i), {\cal{E}}(Y_i), \; i=1,..,k\} \subset {\cal{O}}$.}
\end{itemize}
While more stringent than U3, the condition $\text{U3}'$ is not unreasonable. It grants the measurement device the power to measure two complementary observables for each encoded qubit, and thus to be genuinely quantum. However, the main reason for invoking $\text{U3}'$ is that it removes a substantial amount of technical complication from the proof, while preserving its general structure.}\smallskip

{{\em{Proof of Theorem~\ref{UnivCon} under $U3'$.}} We consider a QCSI where the available initial (magic) states all have an ncHVM description.}

{Now assume that the QCSI scheme is universal for quantum computation. Then, it must be possible to create an encoded Greenberger-Horne-Zeilinger state ${\cal{E}}(|\text{GHZ}\rangle)$, with $|\text{GHZ}\rangle = (|000\rangle + |111\rangle)/\sqrt{2}$, on a subset of the qubits from an initial state ${\cal{E}}(|0\rangle ^{\otimes n})$. Now consider the expectation value 
$$
{\cal{W}}=\langle {\cal{E}}(X_1X_2X_3) -{\cal{E}}(X_1Y_2Y_3) - {\cal{E}}(Y_1X_2Y_3) - {\cal{E}}(Y_1Y_2X_3)\rangle.
$$
${\cal{W}}$ is a contextuality witness derived from Mermin's star \cite{Merm}. Since the observables ${\cal{E}}(X_i)$, $i=1,..,3$, are directly measurable by assumption U$3'$, their product ${\cal{E}}(X_1X_2X_3)$ is inferable, and for any internal state $\nu$ of the ncHVM it holds that
$$
\lambda_\nu({\cal{E}}(X_1X_2X_3)) = \prod_{i=1}^3\lambda_\nu({\cal{E}}(X_i)) .
$$ 
The same holds for the other three measurement contexts $({\cal{E}}(X_1),{\cal{E}}(Y_2),{\cal{E}}(Y_3))$, $({\cal{E}}(Y_1),{\cal{E}}(X_2),{\cal{E}}(Y_3))$, and $({\cal{E}}(Y_1),{\cal{E}}(Y_2),{\cal{E}}(X_3))$. Since for all ncHVM states $\nu$, $\lambda_\nu({\cal{E}}(X_i)), \lambda_\nu({\cal{E}}(X_i)) =\pm 1$, for all states $\rho$ describable by an ncHVM it holds that ${\cal{W}}_\rho \leq 2$, which is the Mermin inequality \cite{Merm2}. ${\cal{W}}_{|\text{GHZ}\rangle}=4$, and the encoded state ${\cal{E}}(|\text{GHZ}\rangle)$ is thus contextual. With Theorem~\ref{Res1}, it cannot be prepared by the given QCSI with a non-zero probability of success. Contradiction.}  

{The indirect assumption is thus wrong. Hence, if the initial (magic) states are non-contextual, the resulting QCSI scheme is not universal. $\Box$}\medskip

{\em{Remark:}} The same conclusion holds when an error $\epsilon$ is allowed in the quantum computation, due to the finite gap between of 2 between ${\cal{W}}_{|\text{GHZ}\rangle}=4$ and ${\cal{W}}_{\rho_{\text{HVM}}}\leq 2$. 

\subsection{Generalized simulation algorithm}\label{GSA}

\begin{Theorem}\label{EffSimGen}
For any QCSI scheme {$(\gamma, {\cal{M}})$}, if (i) the input magic state $\rho_\text{in}$ can be described by a non-contextual HVM with state space ${\cal{S}}$ and value assignments $\lambda_\nu: V \longrightarrow \{\pm1\}$, for all $\nu \in {\cal{S}}$, (ii) this HVM can be efficiently sampled from, and (iii) the value assignments $\lambda_\nu(\textbf{a})$ and the phase convention $\gamma(\textbf{a})$ can be efficiently evaluated  for all $\textbf{a} \in V_{\cal{O}}$, then any resulting QCSI can be efficiently classically simulated. 
\end{Theorem}

Theorem~\ref{EffSimGen} is proved constructively, i.e., by providing a classical simulation algorithm. This algorithm is given in Table~\ref{Algo2}. It is an almost exact copy of the simulation algorithm encountered in Section~\ref{SimulAlg}, and we comment on the resemblance in Section~\ref{AlgoComp}. 

\begin{table}
\begin{tabular}{l}
\textbf{Algorithm 2}\\ \hline \hline
\parbox{8cm}{
\begin{enumerate}
\item{Draw a sample $\nu \in {\cal{S}}$ from the probability distribution $q_{1,\textbf{s}_{\prec 1} = \emptyset}$ describing $\rho_\text{in}$ in the HVM, and set $\nu_1:=\nu$.}
\item{For all the measurements of observables $T_{\textbf{a}_t} \in {\cal{O}}$ comprising the circuit, starting with the first,
\begin{enumerate}
\item{Output the measurement outcome $\lambda_{\nu_t}(\textbf{a}_t) \in \{\pm1\}$ for the observable $T_{\textbf{a}_i}$.}
\item{Flip a fair coin, and update the sample 
\begin{equation}\label{flip}
\nu_t \longrightarrow \nu_{t+1} = \left\{\begin{array}{ll}  \nu_t,&\text{if ``heads''} \\ \nu_t +\textbf{a}_t,& \text{if ``tails'} \end{array} \right. ,
\end{equation}}
\end{enumerate}
until the measurement sequence is complete.}
\item{Repeat until sufficient statistics is gathered.}
\end{enumerate}}\\ \hline 
\end{tabular}
\caption{\label{Algo2}Algorithm 2 for the classical simulation of $n$-qubit QCSI with an ncHVM for the initial state. Addition on the HVM state labels $\nu$ in Eq.~(\ref{flip}) is defined through Eq.~(\ref{Transl}).}
\end{table}

Before we proceed to the proof of Theorem~\ref{EffSimGen}, we  briefly discuss what sampling from conditional probability distributions means for the above algorithm. For any sample $\nu$ drawn in Step 1, while looping through Step 2, a measurement record $\textbf{s}$ is built up. In every iteration $t$ of Step 2, the updated sample $\nu_t$ may be regarded as being drawn from a probability distribution $\tilde{q}_{t,\textbf{s}_{\prec t}}$, conditioned on the previous measurement record $\textbf{s}_{\prec t}$. So the above simulation algorithm definitely samples. The question is whether it samples from the {\em{correct}} distributions, i.e., whether $\tilde{q}_{t,\textbf{s}_{\prec t}}= q_{t,\textbf{s}_{\prec t}}$, for all $t = 1,..,t_\text{max}$ and for all $\textbf{s}$. \medskip

{\em{Proof of Theorem~\ref{EffSimGen}.}} The proof proceeds by demonstrating the correctness and efficiency of the above classical simulation algorithm. 

{\em{Correctness.}} We first show that for each time $t$ and measurement record $\textbf{s}_{\prec t}$, the above classical simulation algorithm (i) produces the correct quantum-mechanical conditional probability $p_t(s_t|\textbf{s}_{\prec t})$ of obtaining the outcome $s_t$ in the measurement of the observable $T_{\textbf{a}_t} \in {\cal{O}}$, and (ii) samples from the correct conditional probability distribution $q_{t,\textbf{s}_{\prec t}}$ of the HVM, which is given by Eq.~(\ref{HVMrelat}). 

The proof is by induction. We assume that at time $t$, the classical simulation algorithm samples from the correct distribution $q_{t,\textbf{s}_{\prec t}}$. 

Re (i): Denote the conditional probabilities produced by the simulation algorithm as $\tilde{p}_t(s_t|\textbf{s}_{\prec t})$. A state $\nu \in {\cal{S}}$ contributes its probability weight $q_{t,\textbf{s}_{\prec t}}(\nu)$ to $\tilde{p}_t(0|\textbf{s}_{\prec t})$ or $\tilde{p}_t(1|\textbf{s}_{\prec t})$ if $\lambda_\nu(T_{\textbf{a}_t}) = + 1$ or $\lambda_\nu(T_{\textbf{a}_t}) = - 1$, respectively. Therefore,
$$
\tilde{p}_t(s_t|\textbf{s}_{\prec t}) = \sum_{\nu \in {\cal{S}}} \delta_{\lambda_\nu(T_{\textbf{a}_t}),(-1)^{s_t}}q_{t,\textbf{s}_{\prec t}}(\nu) = \overline{p}_t(s_t|\textbf{s}_{\prec t}).
$$
The second equality follows by comparison with Eq.~(\ref{HVMp}). Furthermore, $\overline{p}_t(s_t|\textbf{s}_{\prec t})= p_t(s_t|\textbf{s}_{\prec t})$ was already demonstrated in the proof of Theorem~\ref{Res1}. Thus, $\tilde{p}_t(s_t|\textbf{s}_{\prec t}) = p_t(s_t|\textbf{s}_{\prec t})$, as required.

Re (ii): Through the value assignment in step 2(a), an internal state $\nu_t \in {\cal{S}}$ contributes to
$$
\begin{array}{rl}
\tilde{q}_{t+1,(\textbf{s}_{\prec t}, s_t = 0)}, & \text{if } \lambda_{\nu_t}(\textbf{a}_t) = +1,\\
\tilde{q}_{t+1,(\textbf{s}_{\prec t}, s_t = 1)}, & \text{if } \lambda_{\nu_t}(\textbf{a}_t) = -1.
\end{array}
$$
The update rule for Step 2(a) is thus 
$$q_{t, \textbf{s}_{\prec t}}(\tau) \begin{array}{c} s_t\\ \longrightarrow \\ \text{ }\end{array} q^\prime_{t+1,\textbf{s}_{\prec t+1}}(\tau) = q_{t,\textbf{s}_{\prec t}}(\tau) \frac{\delta_{\lambda_\tau(\textbf{a}_t),(-1)^{s_t}}}{p_t(s_t|\textbf{s}_{\prec t})},
$$ 
for all $\tau \in {\cal{S}}$, and  $p_t(s_t|\textbf{s}_{\prec t})$ appears for normalization.

In step 2(b), with Eq.~(\ref{flip}), the update rule is
$$
q^\prime_{t+1,\textbf{s}_{\prec t+1}} \longrightarrow \tilde{q}_{t+1,\textbf{s}_{\prec t+1}} = q^\prime_{t+1,\textbf{s}_{\prec t+1}} * \frac{\delta_{\textbf{0}} +\delta_{\textbf{a}_t}}{2},
$$
where ``$*$'' stands for convolution. Using Eq.~(\ref{Transl}), the resulting expression for  $\tilde{q}_{t+1,\textbf{s}_{\prec t+1}}(\nu)$ matches the expression in Eq.~(\ref{HVMq}), i.e., $\tilde{q}_{t+1,\textbf{s}_{\prec t+1}}(\nu)=q_{t+1,\textbf{s}_{\prec t+1}}(\nu)$, as required. This completes the induction step.

The induction assumption is satisfied for $t=1$, by the first assumption of Theorem~\ref{EffSimGen}. Thus, by induction, the above algorithm samples from the correct conditional outcome probabilities $p(s_t|\textbf{s}_{\prec t})$ for measurement outcomes $s_t$ and from the correct HVM distributions $q_{t,\textbf{s}_{\prec t}}$ describing $\rho_t(\textbf{s}_{\prec t})$, for all times $t$ and all outcome histories  $\textbf{s}$.

{\em{Efficiency.}} The classical preprocessing of removing the unitaries $g \in G$ from the circuit is efficient if the function $\gamma: V \longrightarrow \mathbb{Z}_4$ can be efficiently computed, which holds by assumption. See the proof of Theorem~\ref{Simul}.

Regarding the simulation algorithm itself, the critical step is 2(a), the evaluation of the function $\lambda_\nu$ on some $\textbf{a} \in V$. Again, the efficiency of this function evaluation
 is guaranteed by the assumption of the theorem. $\Box$
 
\subsection{Relation between Algorithms 1 and 2}\label{AlgoComp}

\begin{figure}
\begin{center}
\includegraphics[width=6cm]{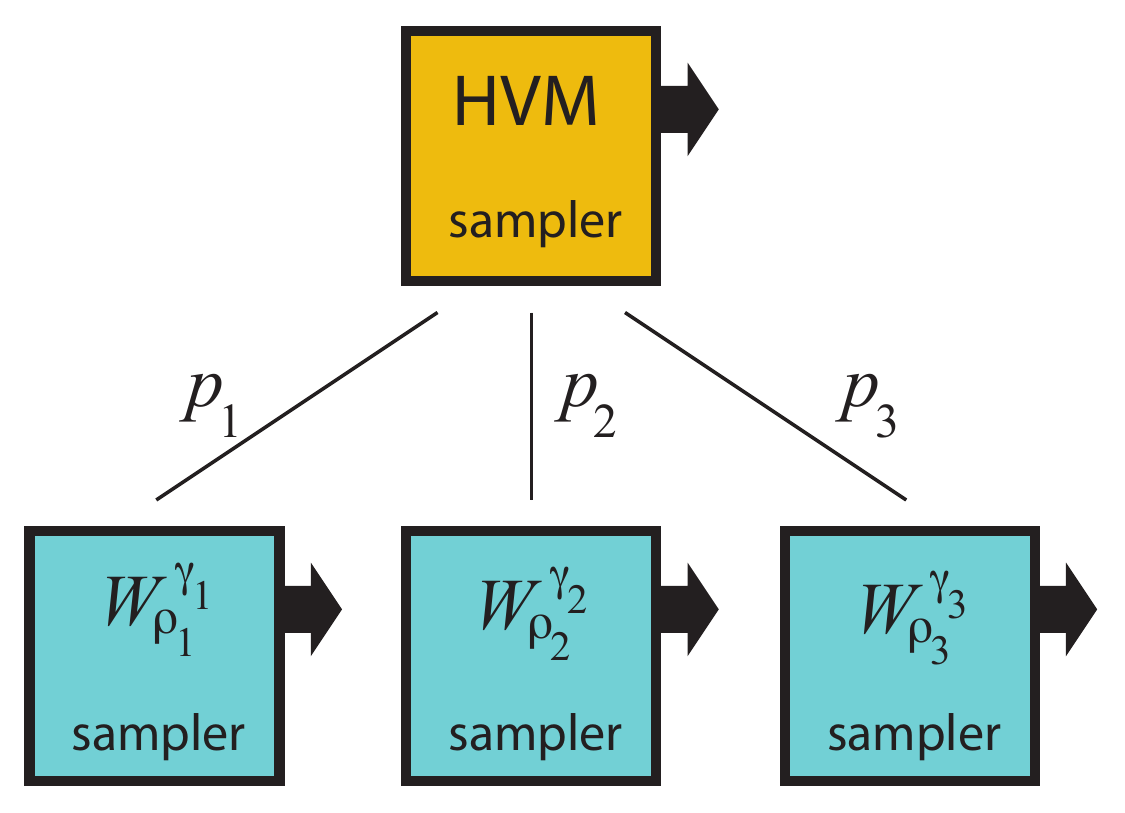}
\caption{\label{SampSim}Relation between Algorithms 1 and 2. Sampling from the probability distribution underlying a non-contextual HVM may be viewed as a two-stage process. Stage 1: Sampling from a probability distribution $\{p_i\}$ over Wigner functions, Stage 2: Sampling from the phase space w.r.t. to the Wigner function chosen in the first stage.}
\end{center}
\end{figure}

Algorithms 1 and 2 are very similar. They only differ in the sampling source they have access to. In this section we explain that Algorithm~2 can be understood as a master algorithm calling Algorithm~1 as a subroutine; See Fig.~\ref{SampSim}. This illustrates that an ncHVM for a set of magic states can be viewed as a probabilistic mixture of non-negative Wigner functions.

By Theorem~\ref{pwnc}, the sampling source for Algorithm 2, based on non-contextual HVMs, is at least as powerful as the sampling source for Algorithm 1, based on non-negative Wigner functions. By the 1-qubit example of Section~\ref{NWNC}, the former source is indeed more powerful. The root of the connection between the two algorithms is that if the initial quantum state $\rho_\text{in}$ can be described by a non-contextual HVM, then it can be represented by an ensemble
$$
{\cal{E}}_{\rho_\text{in}} = \left\{ (p_i,\rho_i)\right\},
$$ 
such that there are Wigner functions $W^{\gamma_i}$ for which (i) $W^{\gamma_i}_{\rho_i}\geq 0$, $\forall i$, and (ii) the measurement of observables from the set ${\cal{O}}$ preserves positivity of the $W^{\gamma_i}$, $\forall i$.

Therefore, Algorithm 2 can be simulated by a master algorithm that merely draws samples $\nu \in {\cal{S}}$ from the non-contextual HVM, and then employs Algorithm 1 as a subroutine for dealing with the samples. This interpretation of Algorithm 2 is developed below.\smallskip

The state space ${\cal{S}}$ of the HVM can be partitioned into orbits $[\nu]$ of $V$,
$$
[\nu] = \{\nu+\textbf{u}, \textbf{u} \in V\} \in {\cal{S}}/V.
$$
Then there exists a special orbit $[0]\in {\cal{S}}/V$ defined by the property that there is a $\tau_{[0]} \in [0]$ for which the value assignment is constant, $\lambda_{\tau_{[0]}}(\cdot) \equiv 1$. With Eq.~(\ref{Transl}) it then follows that $$
\lambda_{\tau_{[0]}+\textbf{u}}(\textbf{a}) = (-1)^{[\textbf{u}, \textbf{a}]},\;\forall \textbf{a} \in V.
$$
Comparing with Eq.~(\ref{ValAss}), we find that the above value assignment  $\lambda_{\tau_{[0]}+\textbf{u}}(\textbf{a})$ agrees with the value assignment made by a positive Wigner function Eq.~(\ref{Adef}) considered as an HVM, if we identify, for all $\textbf{u} \in V$,
$$
(\tau_{[0]}+\textbf{u}) \in [0]  \longleftrightarrow \textbf{u} \in V.
$$
This motivates the definition of a quantum state $\rho_{[0]}$ associated with the orbit $[0]$, via
\begin{equation}\label{Coset0rhoDef}
W_{\rho_{[0]}}(\textbf{u}) :=\frac{q(\tau_{[0]}+\textbf{u})}{p_{[0]}},\; \forall \textbf{u} \in V,
\end{equation}
where $p_{[0]}=\sum_{\textbf{u} \in V} q(\tau_{[0]}+\textbf{u})$ to ensure proper normalization. The state $\rho_{[0]}(\textbf{u})$ is not required to be a valid quantum state, i.e., to be positive semi-definite. The only requirement is a non-negative Wigner function, which is satisfied by definition. The fact that classical sampling algorithms can handle states which have a positive Wigner function but are not proper quantum states is familiar from the qudit case \cite{NegWi}.

In analogy with Eq.~(\ref{Coset0rhoDef}), we will now define states $\rho_{[\nu]}$ for all orbits $[\nu]\in {\cal{S}}/V$. The state $\rho_{[0]}$ and its cousins will then be used in the interpretation of Algorithm 2. 

For any $[\nu] \in {\cal{S}}/V$, pick a $\tau_{[\nu]} \in [\nu]$ and define
\begin{equation}\label{Tdef2}
T^{\gamma_{[\nu]}}_\textbf{a}: = \lambda^{-1}_{\tau_{[\nu]}}(\textbf{a}) T_\textbf{a},\; \forall \textbf{a} \in V,
\end{equation}
where on the r.h.s. $T_\textbf{a} = T^{\gamma}_\textbf{a}$, as defined in Eq.~(\ref{Tdef}). Denoting $\lambda_{\tau_{[\nu]}}(\textbf{a})=(-1)^{s_{[\nu]}(\textbf{a})}$, for all $\textbf{a} \in V$, we thus have the relation
\begin{equation}\label{GammaShift}
\gamma_{[\nu]} \equiv \gamma + 2 s_{[\nu]} \mod 4.
\end{equation}
From the above definition of $T^{\gamma_{[\nu]}}_\textbf{a}$,  $\lambda_{\tau_{[\nu]}}\!\left(T^{\gamma_{[\nu]}}_\textbf{a}\right)=1$, for all $\textbf{a} \in V$. We can thus reproduce for any orbit $[\nu]$ the previous argument for $[0]$. First, with Eq.~(\ref{Transl}),
\begin{equation}\label{CommVal}
\lambda_{\tau_{[\nu]}+\textbf{u}}\!\left(T^{\gamma_{[\nu]}}_\textbf{a}\right)=(-1)^{[\textbf{a},\textbf{u}]},\; \forall \textbf{a},\textbf{u} \in V.
\end{equation}
Again by comparison with Eq.~(\ref{ValAss}), the value assignments by the HVM and by the Wigner function $W^{\gamma_{[\nu]}}$ match if we identify, for all $\textbf{u} \in V$,
\begin{equation}\label{Ident}
(\tau_{[\nu]}+\textbf{u}) \in [\nu]  \longleftrightarrow \textbf{u} \in V.
\end{equation}
A state $\rho_{[\nu]}$ associated with any orbit $[\nu]\in {\cal{S}}/V$ can now be defined, via
\begin{equation}\label{CosetRhoDef}
W^{\gamma_{[\nu]}}_{\rho_{[\nu]}}(\textbf{u}) :=\frac{q(\tau_{[\nu]}+\textbf{u})}{p_{[\nu]}},\; \forall \textbf{u} \in V.
\end{equation}
Therein, $p_{[\nu]}=\sum_{\textbf{u} \in V} q(\tau_{[\nu]}+\textbf{u})$.  As before with $\rho_{[0]}(\textbf{u})$, the state $\rho_{[\nu]}(\textbf{u})$ is not required to be positive semi-definite.

{\em{Remarks:}} (i) For each $[\nu]\in {\cal{S}}/V$,  the choice of the representative $\tau_{[\nu]}$ in Eq.~(\ref{CosetRhoDef}) is arbitrary. Different choices lead to different $\gamma_{[\nu]}$, which are, however, related in a simple way. Namely, the corresponding Wigner functions differ only by translation. By contrast, the Wigner functions $W^{\gamma_{[\nu]}}$ and $W^{\gamma_{[\nu']}}$, for any $[\nu']\neq [\nu]$, are not equivalent under translation.

(ii) We note that multiple Wigner functions have previously been discussed in relation to QCSI \cite{Galv},\cite{Galv2}. Therein, a quantum state is considered classical if {\em{all}} its Wigner functions are positive. Our viewpoint is the opposite. For a state to be considered classical, not even a single one of its Wigner functions has to be positive. \smallskip

The states $\rho_{[\nu]}$, defined in Eq.~(\ref{CosetRhoDef}) have the following relation to the state $\rho$ of the quantum register.
\begin{Lemma}\label{StateRel}
For any QCSI scheme $(\gamma, {\cal{M}})$, if the state $\rho$ of the quantum register has a non-contextual HVM description, then the states $\rho_{[\nu]}$ provide  an ensemble representation ${\cal{E}}_\rho = \left\{(p_{[\nu]},\rho_{[\nu]}), [\nu]\in {\cal{S}}/V \right\}$  of $\rho$, i.e.,
\begin{equation}\label{ensemble}
\rho= \sum_{[\nu] \in {\cal{S}}/V} p_{[\nu]} \rho_{[\nu]}.
\end{equation}
\end{Lemma}
When relating the classical simulation Algorithms 1 and 2, we apply Lemma~\ref{StateRel} in particular to the input state $\rho_\text{in}$ of the computation, i.e., magic state.
 
{\em{Proof of Lemma~\ref{StateRel}.}} Since the $T_\textbf{a}$, $\textbf{a} \in V$, form a basis of Hermitian operators on $n$ qubits, it suffices to show that 
$\langle T_\textbf{a} \rangle_{\rho_\text{in}} = \langle T_\textbf{a} \rangle_{\sum_{[\nu]}p_{[\nu]}\rho_{[\nu]}}$, for all $\textbf{a} \in V$.
$$
\begin{array}{rrl}
\multicolumn{2}{r}{\langle T_\textbf{a} \rangle_{\sum_{[\nu]}p_{[\nu]}\rho_{[\nu]}} }&= \displaystyle{\sum_{[\nu]} p_{[\nu]} \langle T_\textbf{a} \rangle_{\rho_{[\nu]}}}\vspace{1mm}\\
 &=& \displaystyle{\sum_{[\nu]} p_{[\nu]} \lambda_{\tau_{[\nu]}}(\textbf{a}) \langle T_\textbf{a}^{\gamma_{[\nu]}} \rangle_{\rho_{[\nu]}}}\vspace{1mm}\\
&=& \displaystyle{\sum_{[\nu]} p_{[\nu]} \lambda_{\tau_{[\nu]}}(\textbf{a}) 2^n \!\!\sum_{\textbf{u} \in V} W^{\gamma_{[\nu]}}_{\rho_{[\nu]}}(\textbf{u})   W^{\gamma_{[\nu]}}_{T_\textbf{a}^{\gamma_{[\nu]}}} (\textbf{u})} \vspace{1mm}\\
&=& \displaystyle{\sum_{[\nu]}  \lambda_{\tau_{[\nu]}}(\textbf{a}) \!\sum_{\textbf{u} \in V} q(\tau_{[\nu]}+\textbf{u}) (-1)^{[\textbf{a},\textbf{u}]} } \vspace{1mm}\\
&=& \displaystyle{\sum_{\nu \in {\cal{S}}}    q(\nu) \lambda_{\tau_{[\nu]}}(\textbf{a}) \lambda_\nu\left(T_\textbf{a}^{\gamma_{[\nu]}}\right) } \vspace{1mm}\\
&=& \displaystyle{\sum_{\nu \in {\cal{S}}}    q(\nu) \lambda_\nu\left(T_\textbf{a}\right) } \vspace{1mm}\\
&=& \displaystyle{\langle T_\textbf{a} \rangle_{\rho_\text{in}}},
 \end{array}
$$ 
as required. We used Eq.~(\ref{Tdef2}) in line 2 above, Eq.~(\ref{CosetRhoDef})  in line 4, Eq.~(\ref{CommVal}) in line 5, and Eq.~(\ref{Tdef2}) in line 6. $\Box$\smallskip

With Lemma~\ref{StateRel}, we can now re-interpret the sampling from the HVM as the following two-stage process. In the first stage, equivalence classes $[\nu]\in{\cal{S}}/V$ are sampled from, according to the probabilities $\{p_{[\nu]}\}$. In the second stage, given a particular class $[\nu]$, the phase space $V$ is sampled from, according to the conditional probability distribution $q|_{[\nu]}/p_{[\nu]}$.  The conditional probability distributions $q|_{[\nu]}/p_{[\nu]}$ over $V$ are regarded as Wigner functions $W^{\gamma_{[\nu]}}_{\rho_{[\nu]}}$ of states $\rho_{[\nu]}$ associated with the orbits  $[\nu]$, cf. Eq.~(\ref{CosetRhoDef}). See Fig.~\ref{SampSim} for illustration.

Algorithm 2 can thus be simulated by a master algorithm calling Algorithm 1 as a subroutine, as follows. Step 1: A sample $\nu \in {\cal{S}}$ is drawn. This sample is converted into the  into the pair $([\nu]\in {\cal{S}},\textbf{u}\in V)$, such that $\nu = \tau_{[\nu]}+\textbf{u}$. Step 2: Algorithm 1 is called, with the sample $\textbf{u}_1:=\textbf{u}$ being handed over.
 
The orbit  $[\nu]$ has no influence on how Algorithm 1 runs, but needs to be taken into account when the simulated measurement outcomes  are returned. Namely, Algorithm 1 returns the values for $T^{\gamma_{[\nu]}}_{\textbf{a}_t}$, not for the $T_{\textbf{a}_t}$ with the standard phase convention $\gamma$. A conversion of those values is thus necessary, which proceeds by Eq.~(\ref{Tdef2}).

There is one more item to check: To employ sampling from Wigner functions $W^{\gamma_{[\nu]}}$ as a subroutine, the measurement of observables which leave positive Wigner functions $W$ positive must also leave all Wigner functions $W^{\gamma_{[\nu]}}$ positive. Denote by ${\cal{O}}_{[\nu]}$ the non-extendable set of directly measurable observables w.r.t. the phase convention $\gamma_{[\nu]}$ (i.e., the set of Pauli observables whose measurement preserves non-negativity of the Wigner function $W^{\gamma_{[\nu]}}$). Then, we have the following result.
\begin{Lemma}\label{SameO}
For all $[\nu]\in {\cal{S}}/V$, it holds that ${\cal{O}}_{[\nu]}={\cal{O}}$.
\end{Lemma}

{\em{Proof of Lemma~\ref{SameO}.}} By Eq.~(\ref{3T}), for any phase convention $\gamma$ it holds that
$$
\beta(\textbf{a},\textbf{b}) = \gamma(\textbf{a})+\gamma(\textbf{b})-\gamma(\textbf{a}+\textbf{b})+2\textbf{a}_X\textbf{b}_Z \mod 4.
$$
Then, by Eq.~(\ref{GammaShift}), the function $\beta$ based on a specific $\gamma$ and the functions $\beta_{[\nu]}$ based on the corresponding $\gamma_{[\nu]}$ are related via
$$
\beta_{[\nu]}(\textbf{a},\textbf{b}) = \beta(\textbf{a},\textbf{b}) + 2\left(s_{[\nu]}(\textbf{a})+s_{[\nu]}(\textbf{b})-s_{[\nu]}(\textbf{a}+\textbf{b}) \right),
$$
where the addition is again mod 4. Now assume that $T_\textbf{a} \in {\cal{O}}$ and that $[\textbf{a},\textbf{b}]=0$. Then, $\{T_\textbf{a},T_\textbf{b}\}$ is a jointly measurable set of observables, and thus, by Property (ii) of Def.~\ref{HVM1},  $s_{[\nu]}(\textbf{a})+s_{[\nu]}(\textbf{b})-s_{[\nu]}(\textbf{a}+\textbf{b}) \mod 2=0$. Hence,
$$
\beta_{[\nu]}(\textbf{a},\textbf{b}) = \beta(\textbf{a},\textbf{b}),\; \forall\, [\nu] \in {\cal{S}}/V,
$$
for all above pairs $\textbf{a}\in V_{\cal{O}}$, $\textbf{b} \in V$. Thus, by Lemma~\ref{InO}, the measurement of $T_\textbf{a}$ preserves positivity of the Wigner function $W^\gamma$ if and only if it preserves positivity of the Wigner function $W^{\gamma_{[\nu]}}$, for any $[\nu] \in {\cal{S}}/V$. $\Box$

This concludes the discussion of the relation between the classical simulation Algorithms 1 and 2. We have seen that if a given magic state can be described by an ncHVM, then this ncHVM can be viewed as the probabilistic combination of many non-negative Wigner functions, each compatible with the same set ${\cal{O}}$ of measurable observables.
 
\section{A qubit scheme of QCSI with matching Wigner function}\label{qubitQCSI}

Four of the five theorems in the preceding sections begin with ``For any QCSI scheme $(\gamma, {\cal{M}})...$''. We must thus ask: Are there any such schemes at all?---This is the case for any number of qubits, as we now show by example.

\subsection{Definition of the Wigner function}

In this section we focus on the properties of a single Wigner function. We make the choice
\begin{equation}\label{GammaDef}
\gamma_0(\textbf{a}) = \textbf{a}_Z \cdot \textbf{a}_X \!\!\! \mod 4,
\end{equation}
which has the important and rare consequence that the corresponding Wigner function factorizes,  $W_{\rho \otimes \sigma} = W_\rho \cdot W_\sigma$ for all states $\rho$, $\sigma$. 
{In fact, the factorization property already holds on the level of the Heisenberg-Weyl operators Eq.~(\ref{Tdef}),
\begin{equation}\label{Tfact}
T_\textbf{a}\otimes T_\textbf{b} = T_{\textbf{a} + \textbf{b}}.
\end{equation}}

\subsection{The set ${\cal{O}}$  of directly measurable observables}\label{Omod4}

\begin{Lemma}\label{Mset}
For  $\gamma_0(\textbf{a}) = \textbf{a}_Z \cdot \textbf{a}_X \mod 4$, the set ${\cal{O}}$ of directly measurable observables is $${\cal{O}} =\{\pm X_i,\pm Y_i,\pm Z_i|, i =1,..,n\}.$$
\end{Lemma}
This means first of all that the corresponding Wigner function $W^{\gamma_0}$ {\em{has}} a corresponding QCSI scheme, and, perhaps surprisingly, spatial locality plays a role in it. Below, we first prove Lemma~\ref{Mset}, and then flesh out the QCSI scheme corresponding to $W^{\gamma_0}$.\smallskip

{
{\em{Proof of Lemma~\ref{Mset}.}} We first show that the set  ${\cal{O}} =\{X_i,Y_i,Z_i|, i =1,..,n\}$ satisfies the defining conditions Eqs.~(\ref{C1}), (\ref{C2}). Consider two commuting Pauli observables $T_\textbf{b}$, $T_\textbf{c}$ such that $T_\textbf{b}$ is local to qubit $k$, and $T_\textbf{c}$ is written as $T_\textbf{c} = T_{\textbf{c}'+\textbf{c}''}= T_{\textbf{c}'}\otimes T_{\textbf{c}''}$, where $T_{\textbf{c}'}$ acts nontrivially only on qubit $k$, and $T_{\textbf{c}''}$ acts nontrivially only on the complement of qubit $k$. Then,
$$
\begin{array}{rcl}
T_\textbf{b}T_\textbf{c} &=& T_{\textbf{b}} T_{\textbf{c}'}\otimes T_{\textbf{c}''}\\
 &=& i^{-\beta(\textbf{b}, \textbf{c}')}T_{\textbf{b}+\textbf{c}'}  \otimes T_{\textbf{c}''}\\
 &=& i^{-\beta(\textbf{b}, \textbf{c}')}T_{(\textbf{b}+\textbf{c}')+ \textbf{c}''}\\
 &=& i^{-\beta(\textbf{b}, \textbf{c}')}T_{\textbf{b}+\textbf{c}}.
 \end{array}
$$
Therein, in lines 1 and 3 we used the property Eq.~(\ref{Tfact}).}

Since $T_\textbf{b}$ and $T_\textbf{c}$ are commuting, $\beta(\textbf{b}, \textbf{c}') \in \{0,2\}$. Since $T_\textbf{b}$ is local, by going through all the cases of local Pauli operators we find that $\beta(\textbf{b}, \textbf{c}') \in \{0,\pm1\}$. Thus, $\beta(\textbf{b}, \textbf{c}') =0$ is the only choice that satisfies both constraints. Therefore,
\begin{equation}\label{C1pr}
T_{\textbf{b}+\textbf{c}} = T_\textbf{b}T_\textbf{c},
\end{equation}
whenever $[T_\textbf{b}, T_\textbf{c}]=0$, and  $T_\textbf{b}$ is local. Condition~(\ref{C1}) is thus satisfied. 

Next, since every multi-local Pauli operator can be written as a tensor product of local Pauli operators, and the local Pauli operators in such an expansion trivially commute, it follows that $V_M = V$, as required by condition~(\ref{C2}). We have thus shown that $\{X_i,Y_i,Z_i|, i =1,..,n\}$ is a possible set ${\cal{O}} $.

It remains to prove that the above ${\cal{O}}$ is maximal, i.e., that ${\cal{O}}$ cannot contain any additional observable without violating the condition Eq.~(\ref{C1}). 
For $n=1$ this is clear, and we only need to discuss the case of $n\geq 2$. To this end, consider the two-local Pauli operators, beginning with $Y\otimes Y$.
$$
\includegraphics[width=3.9cm]{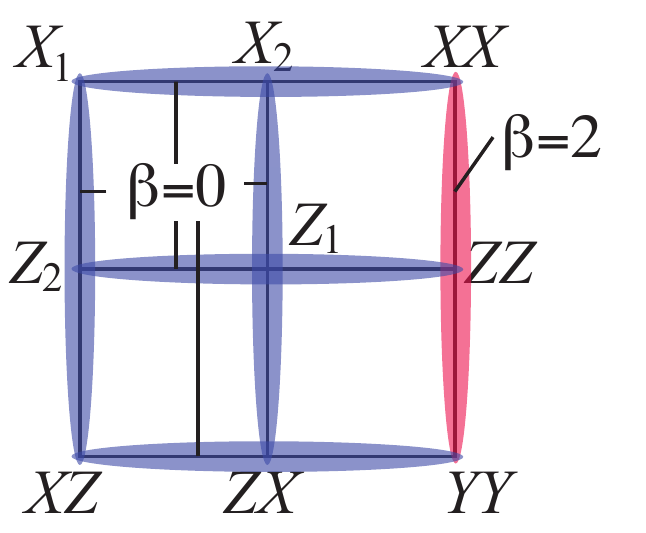}
$$
$Y\otimes Y$ is part of a context with $\beta=2\neq 0 \mod 4$. Therefore, with Lemma~\ref{InO}, $Y\otimes Y \not \in {\cal{O}}$. 

Now, conjugate the observables in the above diagram  under a local Clifford unitary, and readjust the signs such that only observables $T_\textbf{a}$ appear. In this way, any two-local Pauli observable can appear in the bottom left corner of the diagram. We observe that the four observables in the top left corner of the diagram will remain local under such a transformation. As we have shown, all local Pauli operators $T_\textbf{b}$ satisfy Eq.~(\ref{C1pr}) for all commuting $T_\textbf{c}$. Hence, (i) The  four $\beta$s involving local observables remain $\beta=0$. (ii) The six $\beta$ appearing in the square must sum to $2 \mod 4$, as per Mermin's argument. Combining these two facts, we find that the two $\beta$ involving the observable in the bottom right corner of the diagram cannot simultaneously be zero. Hence this observable cannot be in ${\cal{O}}$. Thus, no two-local Pauli observable is in ${\cal{O}}$. 

Next, consider a Pauli observable $T_\textbf{b}$ with a support of size greater than 2. Be $J$ a set of two sites in the support of  $T_\textbf{b}$, $J=\{j,k\} \subset \text{supp}(T_\textbf{b})$, and denote by $T_{\textbf{b}'}$ the restriction of $T_\textbf{b}$ to $J$, and by $T_{\textbf{b}''}$ the restriction of $T_\textbf{b}$ to the complement of $J$. Then, with Eq.~(\ref{Tfact}), $T_\textbf{b} = T_{\textbf{b}'+\textbf{b}''}=   T_{\textbf{b}'} \otimes T_{\textbf{b}''}$.  Now consider a second Pauli operator $T_\textbf{c}$ that commutes with $T_\textbf{b}$ and has support on $J$ only. Then, using the property Eq.~(\ref{Tfact}),
$$
\begin{array}{rcl}
T_\textbf{b}T_\textbf{c} &=& T_{\textbf{b}''} \otimes T_{\textbf{b}'} T_{\textbf{c}}\\
&=& i^{\beta(\textbf{b}',\textbf{c})}T_{\textbf{b}''} \otimes T_{\textbf{b}'+\textbf{c}}\\
&=& i^{\beta(\textbf{b}',\textbf{c})}T_{\textbf{b}'' + (\textbf{b}'+\textbf{c})}\\
&=& i^{\beta(\textbf{b}',\textbf{c})}T_{\textbf{b}+\textbf{c}}.
\end{array}
$$
By the previous argument for two-local operators, for any $T_\textbf{b}$ with support on two or more qubits, a commuting two-local Pauli operator $T_\textbf{c}$ can be found such that $\beta(\textbf{b}',\textbf{c})=2$. Then, with Lemma~\ref{InO}, $\pm T_\textbf{b} \not \in {\cal{O}}$. $\Box$\medskip

From Eq.~(\ref{Om}) it follows that the set $\Omega$ of free states are tensor products of one-qubit stabilizer states. The group of free unitary gates therefore is the local Clifford gates.

\subsection{Magic states and universality}\label{MU}

From the perspective of computational universality of QCSI, all we don't know yet is what the magic states are. Since all  state preparations and measurements are local in the present situation, any entanglement needed in the computation must be brought in by the magic state. That is, there is only one big entangled magic state. Factors of tensor product states cannot be coupled by the free operations. 

\begin{figure}
\begin{center}
\includegraphics[width=8.5cm]{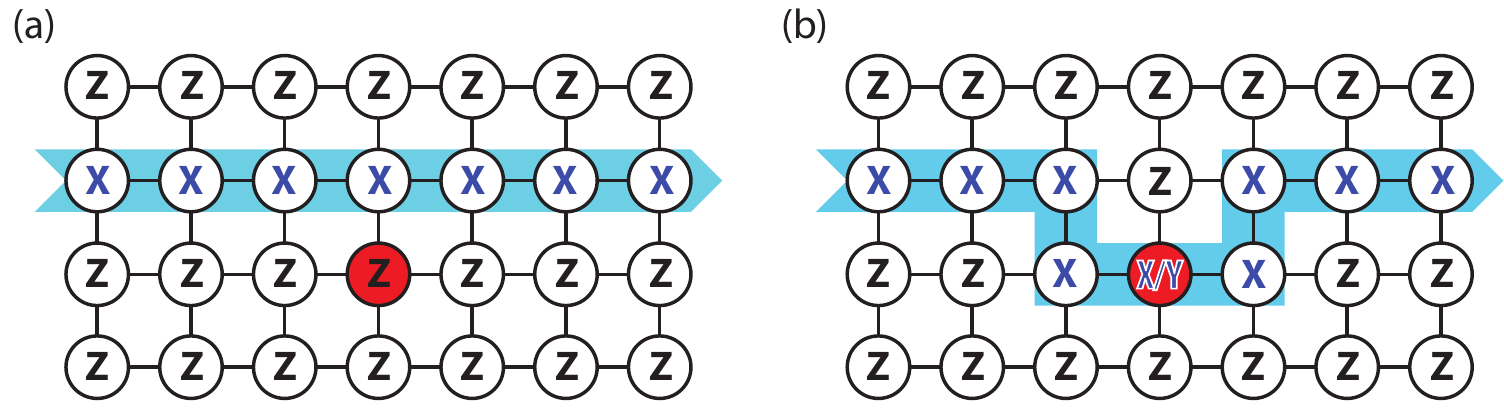}
\caption{\label{cluster} QCSI with modified cluster state of Eqs.~(\ref{KX}), (\ref{KA}) as magic state, which is subjected to measurements of local Pauli operators $X_i$, $Y_j$, $Z_k$, for all $i, j\neq i, k \neq i,j  \in {\cal{V}}$. The role of the $Z$-measurements is to cut out of the plane a web corresponding to some layout of a quantum circuit, and the $X$-measurements drive the MBQC-simulation of this circuit \cite{MBQC}. The qubit in ${\cal{R}}$ is displayed in red. By ``re-routing'' a wire piece, one may choose between implementing and not implementing a non-Clifford gate. (a) Identity operation on the logical state space, (b) Logical gate $e^{i\pi/4\, Z}$.}
\end{center}
\end{figure}

{In fact, one possibility is} to use as magic state a slightly modified cluster state. We consider a set of qubits located on the vertices of a square lattice graph. We denote the set of its sites by ${\cal{V}}$ and its adjacency matrix by $\Gamma$. We single out a subset ${\cal{R}} \subset {\cal{V}}$ of sites which are sufficiently sparse.  Denote by $A$ the observable $\frac{X+Y}{\sqrt{2}}$. With those definitions the resource state $|\Psi\rangle$ is the unique joint eigenstate with eigenvalue 1 of the stabilizer operators
\begin{subequations}
\begin{align}\label{KX}
K^X_a := X_a \bigotimes_{b\in {\cal{V}}}{Z_b}^{\Gamma_{ab}}, \;\;&\text{if } a \in {\cal{V}}\backslash {\cal{R}},\\
\label{KA}
K^A_a := A_a \bigotimes_{b\in {\cal{V}}}{Z_b}^{\Gamma_{ab}}, \;\;&\text{if } a \in {\cal{R}}.
\end{align}
\end{subequations}
That this leads to universal quantum computation is easily shown by standard arguments pertaining to measurement-based quantum computation (MBQC). See Fig.~\ref{cluster} for illustration. 

While being a valid scheme of QCSI, this is also MBQC. The distinction between MBQC and QCSI is thus blurred. By various equivalence transformations, we can make the above computational scheme look more like the known QCSI schemes, or more like standard MBQC.\smallskip 

{\em{Equivalent scheme 1.}} In all QCSI schemes worked out to date \cite{Galv}, \cite{NegWi}, \cite{Howard}, \cite{ReWi}, the magic states are local to single or at most 2 particles.  Although this is by no means necessary, we are used to those states being injected into the computation one by one. If desired, we may convert the above computational scheme into such a form, by conjugating it---the resource state, the measurable observables in ${\cal{O}}$, and the Wigner function $W^0$---under the unitary transformation
\begin{equation}\label{Uising}
U_\text{Ising} = \prod_{i,j\in{\cal{V}}}\left(\Lambda(Z)_{i,j}\right)^{\Gamma_{ij}}.
\end{equation}
In this way, we arrive at the following equivalent computational scheme. The resource state $|\Psi\rangle$ is converted into a tensor product state of individual qubits being in the state $|+\rangle_i$, defined by $X|+\rangle = |+\rangle$, for $i \in {\cal{V}}\backslash {\cal{R}}$, and $|A\rangle_j$, defined by $A|A\rangle=|A\rangle$, for $j\in {\cal{R}}$. The new magic states are thus the local states $|A\rangle_j$.

The new set ${\cal{O}}_1$ of directly measure observables is ${\cal{O}}_1=\{K^X_a, K^Y_a, Z_a,\, a=1,..,n\}$, where
$
K^Y_a = Y_a \bigotimes_{b\in {\cal{V}}}{Z_b}^{\Gamma_{ab}}
$.\smallskip

{\em{Equivalent scheme 2.}} We note in Eq.~(\ref{KA}) that stabilizer operators $K^A$ of the magic state $|\Psi\rangle$ are not exactly stabilizer operators of cluster states. Therefore, we may apply the equivalence transformation
$$
U_\text{loc} = \bigotimes_{j\in {\cal{R}}}e^{-i\pi/4\, Z_j},
$$ 
and obtain as the new magic state the standard cluster state. The new measurable observables are
$$
{\cal{O}}_2=\{X_i,Y_i, A_j,A'_j,Z_k|\,a\in {\cal{V}}\backslash {\cal{R}}, j \in {\cal{R}}, k=1..n\},
$$ 
where $A'=(X-Y)/\sqrt{2}$. We note that the measurable observables which are not $Z$s are of the form
$$
O_i = \cos \phi_i\, X_i \pm \sin \phi_i \, Y_i,
$$
as standard in MBQC \cite{MBQC}. A minor deviation from the standard remains. Namely, for each site $i$, only a single setting out of two is available for the measurement angle, either $\phi_i=0$ or $\phi_i =\pi/4$. In standard MBQC, any angle $\phi_i \in [0,\pi/2]$ may be chosen. However, the present restriction does not affect computational universality.

\section{Conclusion}\label{Concl}

We have shown that for all schemes of quantum computation with magic states on qubits that satisfy the condition (P1), contextuality and Wigner function negativity are necessary resources for quantum computational universality and quantum speedup. This extends the earlier results  \cite{Howard},\cite{NegWi} for qudits to qubits. Key to our construction is the additional condition (P1) imposed on QCSI schemes, which removes the phenomenon of state-independent contextuality from the free operations.

Despite the interchangeability of ``contextuality'' and ``Wigner function negativity'' in the above statement about computational resources, the results on contextuality are stronger. They imply their counterparts on the negativity of Wigner functions, but are not implied by them. This arises because the equivalence between the existence of a non-contextual hidden variable model and non-negativity of the Wigner function---which characterized the qudit case \cite{Howard}, \cite{Nicol}---no longer holds for qubits. Magic states can be described by an ncHVM but nonetheless have a negative Wigner function. 

A related matter is the preservation of positivity under free QCSI operations. For any given QCSI scheme satisfying (P1), we find that the existence of an ncHVM for the state of the quantum register is preserved under all free operations. The non-negativity of Wigner functions is only preserved under free measurements, not under all free unitaries. The amount of negativity introduced by the free unitaries can be very large as measured by the sum negativity \cite{EmRes}, without compromising efficient classical simulability.

We conclude with two questions.
\begin{itemize}
\item{We have shown that QCSI schemes satisfying the conditions (P1) and (P2) exist for any number $n$ of qubits, but  we have not classified them. From initial numerical studies we know: (i) For $n=2$ there are plenty of QCSI schemes---as specified by the function $\gamma$ in Eq.~(\ref{Tdef})---that satisfy both (P1) and (P2). In particular, there are at least two distinct classes under Clifford equivalence. This tells us that there are solutions fundamentally different from Eq.~(\ref{GammaDef}). (ii) For $n\geq3$, the solutions of (P1), (P2) for $\gamma$ are very sparse (we didn't find any in $10^4$ random trials). Yet, by Lemma~\ref{Mset}  we know that solutions exist for any $n$. Can they be classified?}
\item{Determining the cost of classically simulating universal quantum computing is fascinating area of research \cite{Bartl},\cite{Nest},\cite{BSS},\cite{HC}. Can the simulation methods developed here be extended to qubit QCSI schemes that do not satisfy the constraint (P1)?}
\end{itemize}

\medskip

\medskip

{\em{Acknowledgments.}} RR thanks Terry Rudolph and the Horodecki family for discussion. This work has been funded by NSERC (ND, CO and RR), SIQS (JBV), AQuS (JBV), Cifar (ND and RR), and IARPA (RR). RR is Fellow of the Cifar Quantum Information Program.

\appendix

\section{Proof of Theorem~\ref{UnivCon}}\label{UnivCP}

Before we give the proof, we need to set up some additional notation. Namely, we extend the value assignments $\lambda_\nu$. As per Eq.~(\ref{flip}) in the simulation Algorithm 2 of Section~\ref{GSA}, the update of the internal state $\nu$ in the measurement of an observable $T_{\textbf{a}_t}$ invokes a random bit $c_t$, $\nu_t \longrightarrow \nu_{t+1} = \nu_t+c_t \, \textbf{a}_t$. Assume, the measurement sequence representing a given quantum algorithm has a maximum length $t_\text{max}$. Since the values $\{c_t, t = 1, .. , t_\text{max}\}$ are random and uncorrelated with everything, it does not matter whether they are determined at runtime or before the computation, and we may include those values in the ncHVM value assignments $\lambda_\nu$. That is, we extend the internal states $\nu \in {\cal{S}}$ to
$$
\nu^* = (\nu, c_1, c_2,\ldots, c_{t_\mathrm{max}}).
$$
Correspondingly, the set of internal states becomes ${\cal{S}}^*:= {\cal{S}} \times \mathbb{Z}_2^{\times t_\text{max}}$.

A consequence of this definition is that the function $\lambda_\nu$ can now assign a value to a potentially larger set of observables than the inferable ones. Namely, consider a unitary $U$ which can be implemented using free gates, free measurements and ancilla states that can be described by an ncHVM, and an inferable observable $O \in M$. Then, the observable $U^\dagger O U$ can also be measured, namely by first implementing $U$ and then measuring $O$. However, it is not {\em{a priori}} clear that $U^\dagger O U \in M$.  Yet, by the general simulation Algorithm~2, given the initial internal state $\nu^*$, a value $\lambda_{\nu^*}(U^\dagger O U)$ can be assigned to the observable $U^\dagger O U$.

An important point to note is that the above value assignments involving quantum circuits are in general context-dependent. That is, they depend on how the unitary $U$ is implemented as a circuit, and which acilla states are being consumed in the process. Further, for two given observables $O\in M$, $O'$, related via $O' = U^\dagger O U$, the unitary $U$ is not unique, and $\lambda_{\nu^*}(U^\dagger O U)$ may again depend on the choice made. It is this potential contextuality of value assignments involving circuits that complicates the proof of Theorem~\ref{UnivCon}.

Let's illustrate this property in an example. Suppose there is a QCSI setting where all observables $X_i$ can be directly measured but $Y_3$ cannot. Further, assume that the unitary $S_3$, with the property $Y_3 = S_3^\dagger X_3 S_3$, can be implemented using only free operations and ancillas that have an ncHVM description. The circuit for measuring $Y_3$ in this example is to first implement $S_3$ and then measure $X_3$. Now the task is to measure $Y_3$ after first measuring either $X_1$ or $X_2$.  Are the corresponding values $\lambda_{\nu^*}(Y_3)$ assigned by the internal state $\nu^*$ via the simulation Algorithm~2 the same in those two situations?

This is in general not the case. Suppose that $\nu^*$ is such that $c_1=1$. Then, if $X_1$ was measured first, the HVM state after the first measurement is $\nu+ \textbf{a}_{X_1}$. Analogously, if $X_2$ was measured first, the HVM state after the first measurement is $\nu+ \textbf{a}_{X_2}$. With the second measurement in the sequence begins the implementation of the measurement of $Y_3$. By construction, the measurement strategy does not depend on the outcome of the first measurement. Therefore, the second measured observable is the same $T_{\textbf{a}_2}$ in both cases. However, the value assigned to this observable by the internal state $\nu^*$ through Eq.~(\ref{Transl}) differs if $[\textbf{a}_2, \textbf{a}_{X_1}] \neq [\textbf{a}_2, \textbf{a}_{X_2}]$. In this situation, the HVM prediction for the outcome obtained in the second measurement depends on which observable was measured first. Since the choice of observable in the third measurement may depend on the outcome of the second measurement, the two measurement sequences may now run down completely different tracks, and there is no guarantee whatsoever that the value assigned to $Y_3$ will be the same. Hence, the value assignments involving circuits are potentially contextual. \medskip

{{\em{Proof of Theorem~\ref{UnivCon}.}} The outline of the proof is as follows. Suppose the statement is wrong, i.e., it is possible that the initial magic state has a description of an ncHVM and simultaneously gives rise to universal quantum computation. Then, by assumptions U1 and U2, universality implies that an encoded GHZ-state can be prepared. Further, by U2 and U3 it is possible to randomly measure in one of the four contexts ${\cal{E}}(X_1,X_2,X_3)$,  ${\cal{E}}(X_1,Y_2,Y_3)$,  ${\cal{E}}(Y_1,X_2,Y_3)$,  ${\cal{E}}(Y_1,Y_2,X_3)$. In each such measurement, the three outcomes are multiplied and it is verified whether the product matches the corresponding stabilizer eigenvalue of the GHZ state.} 

{We will show below that if the initial (magic) state has an ncHVM description, then the simulation Algorithm 2 will fail to reproduce the quantum prediction {for the stabilizer eigenvalues of the GHZ-state} with a probability $\geq 1/8$. However, Algorithm 2 has been proven to be correct (See Theorem~\ref{EffSimGen}), and therefore computational universality and availability of an ncHVM description for the initial state cannot simultaneously apply.}
 \medskip

{We now describe the measurement procedures for the four measurement contexts. Case (i): ${\cal{E}}(X_1,X_2,X_3)$. The three observables ${\cal{E}}(X_1)$, ${\cal{E}}(X_2)$, ${\cal{E}}(X_3)$ are jointly measured, which is guaranteed to be possible by U3.}

{Case (ii): the three remaining contexts ${\cal{E}}(X_i,Y_j,Y_k)$, $i\neq j \neq k \in \{1,2,3\}$. For each of the three measurement contexts, we apply the following procedure:
\begin{enumerate}
\item{Measure the observable ${\cal{E}}(X_i)$.}
\item{Implement the measurement sequence corresponding to the execution of the encoded gate ${\cal{E}}(S_1\otimes S_2 \otimes S_3)$, giving rise to a free unitary $g\in G$ which is propagated out of the circuit; See Fig.~\ref{Conj}.}
\item{Measure the observables $g^\dagger {\cal{E}}(X_i) g$, for $i=1,..,3$. The outcomes for ${\cal{E}}(Y_1)$, ${\cal{E}}(Y_2)$ and ${\cal{E}}(Y_3)$ are thereby obtained. Among them, the value for ${\cal{E}}(Y_i)$ is discarded.}
\end{enumerate}
Step 1 of this procedure is possible by U3, Step 2 is possible by U2. Regarding Step 3, the unitary $g$ is in general dependent on them measurement outcomes obtained in implementing the circuit for ${\cal{E}}(S_1\otimes S_2 \otimes S_3)$, but in all cases $g$ is free ($g\in G$). Thus, the observables $g^\dagger {\cal{E}}(X_i) g$ are in ${\cal{O}}$ and can be directly measured.}\medskip

{From the perspective of the hidden variable model, for any internal state $\nu^* \in {\cal{S}}^*$, the values $\lambda_{\nu^*}({\cal{E}}(X_i))$, $\lambda_{\nu^*}({\cal{E}}(Y_i))$, $i=1,..,3$ are assigned to the corresponding observables via the simulation Algorithm~2 applied to the above measurement procedure. We now show that these value assignments are non-contextual if $c_1=0$.}

{(i) ${\cal{E}}(X_i)$. Since the ${\cal{E}}(X_i)$ are measured first in every context they appear in, $\lambda_{\nu^*}({\cal{E}}(X_i)) = \lambda_{\nu}({\cal{E}}(X_i))$. They are thus context independent.}

{(ii) ${\cal{E}}(Y_i)$. We show this by the example of ${\cal{E}}(Y_1)$, the other two cases are analogous. ${\cal{E}}(Y_1)$ appears in two measurement contexts, namely  ${\cal{E}}(Y_1,X_2,Y_3)$ and  ${\cal{E}}(Y_1,Y_2,X_3)$. The corresponding measurement procedure differ only in one respect. In the former context ${\cal{E}}(X_2)$ is measured in the first step, and in the latter context ${\cal{E}}(X_3)$ is measured in the fist step. Now, by the assumption of $c_1=0$, the internal state $\nu$ remains unchanged under this first measurement in both cases, and in particular remains the same for both cases. After the first step, the procedures are thus identical, and the same value $\lambda_{\nu^*}({\cal{E}}(Y_1))$ is assigned by the HVM in either context.}

{Thus, if $c_1=0$, all values $\lambda_{\nu^*}({\cal{E}}(X_i))$, $\lambda_{\nu^*}({\cal{E}}(Y_i))$, $i=1,..,3$ assigned by the HVM are thus non-contextual, as claimed. By Mermin's argument, such non-contextual value assignments fail to reproduce the respective stabilizer eigenvalue of the encoded GHZ-state in at least one of the contexts. Thus, if the measurement contexts are randomly chosen, if $c_1=0$ then the HVM fails to match the quantum prediction with a probability of $\geq 1/4$.}

{Further, for any internal state $\nu\in {\cal{S}}$, the conditional probability for $c_1=0$ in $\nu^*$ given $\nu$ equals 1/2 (recall that $c_1$ is the result of an unbiassed coin flip). Therefore, the HVM fails to match the quantum prediction with a probability $\geq 1/8$.} 

{Finally, universality only requires the preparation of an encoded GHZ-state ${\cal{E}}(|\text{GHZ}\rangle)$ up to an arbitrarily small error $\epsilon>0$, and likewise only the approximate realization of the gates ${\cal{E}}(S_i)$, up to the same error $\epsilon$. When taking such errors into account, the HVM prediction remains unchanged. If the initial state can be described by a non-contextual HVM, the measurement record will fail to reproduce the quantum mechanical prediction of the measurement record with a probability $\geq 1/8$.}

{The quantum prediction does now differ from the ideal scenario. With a slightly erroneous encoded GHZ-state ${\cal{E}}(|\text{GHZ}\rangle)$ and slightly erroneous gates ${\cal{E}}(S_i)$, the measurement record will differ from the measurement record for the ideal circuit with a probability $p(\epsilon)$. However, $p(\epsilon)$ can be made arbitrarily small by decreasing $\epsilon$. Thus, even in the presence of a small error $\epsilon$, the quantum prediction can be distinguished from the HVM prediction.
$\Box$}

\end{document}